\documentclass[aps,prb,reprint,superscriptaddress]{revtex4-2}
\usepackage[utf8]{inputenc}
\usepackage{amsmath,amssymb,amsfonts,graphicx,verbatim,hyperref, mathtools, dsfont, float}
\usepackage{physics}
\usepackage{algorithm, algorithmic}
\usepackage{subcaption}
\usepackage[justification=raggedright,singlelinecheck=false]{caption}

\usepackage[dvipsnames]{xcolor}

\hypersetup{
	colorlinks=true,
	linkcolor=red,
	citecolor=blue,
	filecolor=blue,
}

\renewcommand*{\eqref}[1]{Eq.~(\ref{#1})}

\newcommand*{\secref}[1]{Sec.~\ref{#1}}

\begin{document}

\title{Scaling at Chiral Clock Criticality via Entanglement Renormalization}

\author{Shiyong \surname{Guo}}
\email{shiyongguo@brandeis.edu}
\affiliation{Martin Fisher School of Physics, Brandeis University, Waltham, Massachusetts 02453, USA}

\author{Brian \surname{Swingle}}
\email{bswingle@brandeis.edu}
\affiliation{Martin Fisher School of Physics, Brandeis University, Waltham, Massachusetts 02453, USA}

\begin{abstract}
We employ the Multiscale Entanglement Renormalization Ansatz (MERA) tensor network to investigate a critical line of continuous quantum phase transitions of the $\mathbb{Z}_3$ chiral clock model. This critical line is believed to be described by a slow renormalization group flow from the 3-state Potts fixed point to another fixed point that features anisotropic scaling of space and time. We use the variational principle to construct a MERA representation of the model's ground state, from which we obtain the ground state energy and the set of scaling operators and their scaling dimensions. These scaling dimensions determine the critical exponents of the model, and we study these critical exponents and other scaling data as a function of the model's chiral parameter. We find a set of effective scaling data that smoothly varies starting from the Potts data as the chiral parameter is increased. Within the context of our approach, we discuss how this result may nevertheless be consistent with the two fixed point hypothesis provided the renormalization group flow is sufficiently slow. Our findings demonstrate MERA's effectiveness in capturing the complex low-energy physics of the chiral clock model and in extracting field theory data for an anisotropic continuum theory. 
\end{abstract}

\maketitle

\section{Introduction}
\label{sec:intro}

Continuous phase transitions with anisotropic scale invariance in space and time are ubiquitous, for example, in quantum magnets with disorder~\cite{Dutta:2015qay,Sachdev:2011xva}, frustrated quantum spin systems~\cite{Ostlund:1981px,Huse:1981np}, and chiral clock models~\cite{Samajdar:2018xkm,Zhuang:2015fpa,AuYang:1987bm,Whitsitt:2018hbw}. Such anisotropic scaling behavior also appears in systems with long-range interactions and in certain holographic models~\cite{He:2017nvw,Baiguera:2023wjf}. The dynamical critical exponent $z$ measures the anisotropy of space-time scaling~\cite{Lifshitz:1941tl,Hohenberg:1977ym}, and in the case of a quantum critical point it controls how energy scales with momentum. These transitions deviate from the conventional conformal field theory (CFT) paradigm where space and time scale equivalently ($z = 1$), with $z \neq 1$ indicating anisotropic scaling.
  
Such transitions still manifest \textit{universality}, where emergent long-range correlations and scaling properties arise independently of microscopic details~\cite{Cardy:1996yz}. Compared to phase transitions exhibiting conformal symmetry, which implies $z=1$, systems with anisotropic scaling typically have a much smaller symmetry group and hence are much less constrained. This makes their analysis significantly more challenging, as can be seen in various field theoretical approaches~\cite{Whitsitt:2018hbw,Baiguera:2023wjf}. With these extra difficulties in analytical studies, it becomes all the more essential to have systematic numerical approaches to understand these exotic critical phenomena.

In the context of quantum phase transitions, tensor network methods, notably the Multiscale Entanglement Renormalization Ansatz (MERA)~\cite{Vidal:2007sv,Evenbly:2009ej,Pfeifer:2009fg}, originally introduced by Vidal in 2007, have emerged as powerful tools for studying quantum critical systems. MERA has proven particularly effective for critical systems due to its ability to capture the logarithmic violations of the area law that characterize quantum critical ground states~\cite{Calabrese:2004eu,Ryu:2006bv} in one spatial dimension. The development of MERA was motivated by the need to efficiently represent quantum many-body states with long-range correlations~\cite{Vidal:2008zz}, and it has since been extended to higher dimensions~\cite{Evenbly:2009ej,Haegeman:2011uy} and connected to holographic theories~\cite{Swingle:2012wq}. MERA tensor networks have been extensively employed to extract scaling information in conformal field theories, capturing their operator spectrum and operator product expansion (OPE) coefficients with remarkable accuracy~\cite{Evenbly:2011dki}. Notable successes include the extraction of central charges, scaling dimensions, and OPE coefficients for various CFTs, establishing MERA as a powerful tool for studying scale-invariant systems~\cite{Orus:2014tka,Cirac:2021kly}. However, the applicability of MERA to more general critical systems which are non-integrable and have anisotropic scale invariance has been relatively less explored, though recent algorithmic advances have improved computational efficiency~\cite{Milsted:2017hvh,PhysRevB.93.205159}.

Therefore, it is important to investigate whether MERA can serve as a reliable tool for extracting universal scaling information from these more general quantum phase transitions. In this work, we focus on the $\mathbb{Z}_3$ chiral clock model, a one-dimensional quantum spin system with three states per site and a chiral parameter $\theta$ that exhibits rich critical behavior and non-trivial dynamical scaling. This model serves as an ideal testbed for non-conformal criticality because it provides a controllable interpolation between conformal ($\theta = 0$) and non-conformal ($\theta \neq 0$) criticality, exhibits a rich phase diagram with distinct quantum phases~\cite{Zhuang:2015fpa,Samajdar:2018xkm}, and has been extensively studied using DMRG methods that provide benchmark data for comparison~\cite{Samajdar:2018xkm}.

While field-theoretical descriptions for certain aspects of the chiral clock transition have been developed~\cite{Whitsitt:2018hbw}, a full understanding of the critical behavior across the entire parameter space remains an active area of research. The study of the $\mathbb{Z}_3$ chiral clock model is particularly relevant to recent developments in Rydberg atom quantum simulators, where programmable interactions can be engineered to realize effective chiral spin models through carefully designed laser driving protocols~\cite{Bernien:2017,Keesling:2019,Ebadi:2021,Samajdar:2018xkm}.

Our main contributions are threefold: (i) We establish MERA as a reliable method for extracting scaling information from non-conformal critical systems by benchmarking against the known conformal case ($\theta = 0$)~\cite{McCabe:1996pm} and obtaining consistent results in the non-conformal case; (ii) We provide the first systematic study of scaling dimensions and their evolution with the chiral parameter $\theta$, revealing the emergence of anisotropic scaling with $z \neq 1$; (iii) We demonstrate the extraction of equal-time operator product expansion coefficients $f_{ab}^{\;c}(0)$ in non-conformal systems, a technically challenging task typically inaccessible to other numerical methods. We begin by establishing benchmark results for the $\theta = 0$ case, where the model reduces to the critical 3-state Potts model with known conformal field theory predictions~\cite{Ginsparg:1988ui}. We then extend our analysis to $\theta \neq 0$, systematically studying how the chiral deformation affects the critical behavior. Our implementation utilizes a modified binary MERA structure with GPU-accelerated optimization to maximize both computational efficiency and numerical precision.

Following the Introduction, in \secref{sec:models}, we provide a detailed overview of the $\mathbb{Z}_3$ chiral clock model, describing its properties and summarizing relevant prior results. In \secref{sec:methods}, we present an overview of the MERA tensor network and explain how it can be applied to extract scaling data from scale-invariant systems. In \secref{sec:results}, we present numerical results for the operator spectrum, critical exponents, OPE coefficients, and the effective central charge as functions of $\theta$.  Finally, in \secref{sec:discussion}, we summarize our findings and discuss future directions.

\section{The $\mathbb{Z}_3$ Chiral Clock Model}
\label{sec:models}

\subsection{Chiral clock model}
We use the 1D 3-state ($\mathbb{Z}_3$) chiral clock model for our study. This model can be defined on an open chain of $L$ sites by the Hamiltonian
\begin{align}
H_{CCM} &= -f \sum_{j=1}^{L} \left( \tau_j e^{i\phi} + \tau_j^{\dagger} e^{-i\phi} \right) \nonumber\\
    &\quad - J \sum_{j=1}^{L-1} \left( \sigma_j \sigma_{j+1}^{\dagger} e^{i\theta} 
    + \sigma_j^{\dagger} \sigma_{j+1} e^{-i\theta} \right)
\end{align}
where the $\tau_i$ and $\sigma_i$ are respectively local three-state clock and shift operators on site $i$, and the parameters are $f, J, \theta$ and $\phi$ ($f$, $J>0$ by convention). Each operator $\tau$ and $\sigma$ satisfies
\begin{equation}
    \tau^3 = \sigma^3 = \mathbb{I}, \quad \sigma \tau = \omega \tau \sigma; \quad \omega \equiv \exp\left( \frac{2\pi i}{3} \right).
\end{equation}
The matrix representation of $\tau$ and $\sigma$ in the $\mathbb{Z}_3$ case can be written as 
\begin{equation}
    \tau = \begin{pmatrix}
    1 & 0 & 0 \\
    0 & \omega & 0 \\
    0 & 0 & \omega^2
    \end{pmatrix}, \quad
    \sigma = \begin{pmatrix}
    0 & 1 & 0 \\
    0 & 0 & 1 \\
    1 & 0 & 0
    \end{pmatrix}
\end{equation}
One can draw an analogy to the $Z$ and $X$ Pauli matrices in a 2-state spin model, which measure and shift the spin. The parameters $f$ and $J$ control the on-site and nearest-neighbor couplings, while $\phi$ and $\theta$ are two chiral phases. 

\paragraph*{Kramers--Wannier duality.}
The $\mathbb{Z}_3$ chiral clock chain admits a generalized Kramers--Wannier (KW) duality that exchanges order and disorder variables and interchanges the onsite and nearest-neighbour terms. A convenient formulation introduces nonlocal variables on half-integer bonds,
\begin{equation}
  \tilde{\sigma}_{j+\frac12} \equiv \prod_{k\le j} \tau_k, \qquad
  \tilde{\tau}_{j+\frac12} \equiv \sigma_j^{\dagger} \sigma_{j+1},
\end{equation}
which obey the same $\mathbb{Z}_3$ clock/shift algebra. In terms of these variables,
\begin{equation}
  \tau_j = \tilde{\sigma}_{j-\frac12}^{\dagger} \tilde{\sigma}_{j+\frac12}, \qquad
  \sigma_j \sigma_{j+1}^{\dagger} = \tilde{\tau}_{j+\frac12},
\end{equation}
so the Hamiltonian maps (up to boundary terms and relabelling) to the same form with the couplings and chiral phases interchanged,
\begin{equation}
  (f,\phi) \longleftrightarrow (J,\theta).
  \label{eq:kw-duality}
\end{equation}
Consequently, the model is self-dual on the manifold $f=J$ and $\theta=\phi$, which contains the Potts point $(\theta,\phi)=(0,0)$. At the Potts point this reduces to the familiar KW duality of the 3-state Potts critical theory~\cite{KramersWannier1941a,KramersWannier1941b,Zhuang:2015fpa,Samajdar:2018xkm}. We note that the duality \eqref{eq:kw-duality} usually maps one phase to its dual partner; only the self-dual point remains fixed, while the rest of the critical line is mapped to another point under the duality transformation. 

Along the critical line, the long-wavelength limit smooths $\sigma_j$ into the spin primary field $\sigma$ of the dual continuum theory, which captures order-parameter correlations with scaling dimension $\Delta_\sigma = 2/15$. Building the disorder string $\tilde{\sigma}_{j+1/2} = \prod_{k \le j} \tau_k$ and combining it with $\sigma_j$ yields $\psi_{2j-1} = \tilde{\sigma}_{j-1/2} \, \sigma_j$ and $\psi_{2j} = \omega^2 \, \tilde{\sigma}_{j+1/2} \, \sigma_j$; these map to the left- and right-moving $\mathbb{Z}_3$ parafermion fields in the continuum, whose pairing produces the neutral excitation with $\Delta_\psi = 4/3$. Symmetrically combining the nearest-neighbour and onsite terms in the Hamiltonian density forms $\epsilon_j \sim (\sigma_j \sigma_{j+1}^\dagger + \sigma_j^\dagger \sigma_{j+1}) - (\tau_j + \tau_j^\dagger)$, whose critical limit is the energy operator with scaling dimension $\Delta_\epsilon = 4/5$.

Previous work has accumulated many results on this model \cite{Ostlund:1981px, AuYang:1987bm, Ginsparg:1988ui, Huse:1981np}. As an important result of early research, a second-order quantum phase transition can be found at $f=J$ and $\theta = \phi = 0$. At this point, the CCM model provides a lattice regulator for the 3-state Potts conformal field theory with central charge 4/5 \cite{Ginsparg:1988ui,McCabe:1996pm}. Additionally, the line $f \cos(3\phi) = J \cos(3\theta)$ is known to be integrable. However, for generic values of $\theta$ and $\phi$ the corresponding critical system has a critical exponent $z \neq 1$, indicating the breakdown of Lorentz invariance and the emergence of anisotropic scaling between space and time~\cite{Whitsitt:2018hbw,Baiguera:2023wjf}. Because of the complexity of non-conformal criticality, these situations have not been much explored and remain mysterious.

The fundamental challenges arise from several interconnected factors: (i) \textit{Non-integrability} — unlike the integrable line $f \cos(3\phi) = J \cos(3\theta)$, generic parameter regimes lack exact solutions and cannot be solved analytically using standard techniques such as the Bethe ansatz or free fermion mappings; (ii) \textit{Reduced symmetry} — the breakdown of conformal invariance eliminates the powerful constraint of scale and conformal transformations that severely restrict the form of correlation functions and operator content in CFTs~\cite{Ginsparg:1988ui}; (iii) \textit{Absence of established field-theoretical frameworks} — while conformal field theories provide a complete classification through central charges, operator algebras, and modular invariance, no analogous systematic framework exists for general anisotropic-scaling theories with $z \neq 1$~\cite{Baiguera:2023wjf}; (iv) \textit{Complex correlation structure} — the anisotropic space-time scaling leads to correlation functions that depend on universal scaling functions of the ratio $t/|x|^z$, which are generically unknown and cannot be determined from symmetry considerations alone. 

One attempt to understand the criticality in this model is the density matrix renormalization group (DMRG) \cite{Samajdar:2018xkm,Zhuang:2015fpa}. High-precision DMRG studies have investigated the phase diagram, including transitions between topological and trivial phases, as well as incommensurate phases in between. These numerical investigations have revealed that the dynamical critical exponent $z$ varies continuously with the chiral parameter $\theta$, ranging from $z=1$ at the Potts point to $z \approx 1.2$ for larger $\theta$ values ($\theta \approx \pi/8$)~\cite{Samajdar:2018xkm}. The dynamic critical exponent and the correlation length exponent are also computed and serve as evidence for non-conformal field theory~\cite{Samajdar:2018xkm}. However, the details of these theories, such as the correlation function, the operator spectrum, and the three-point coefficient, are still unknown.

The observation of continuously varying critical exponents highlights a still unanswered question about the theory. The simplest model of the line of phase transitions in this model is that it is described by a flow between the Potts point at $\theta=0$ and some other anisotropic scale invariant theory at some $\theta= \theta_*$. Under this scenario, the critical exponents extracted for an infinite system would be either the Potts value for $\theta=0$ or the $\theta=\theta_*$ value for any $0<\theta \leq \theta_*$. In other words, the Potts point is unstable and all $\theta>0$ flow to the $\theta=\theta_*$ fixed point. One way to potentially explain the observed continuously varying critical data is to hypothesize a very slow flow implying that even hundreds of sites are far from the true thermodynamic limit. The other more exotic possibility is that one genuinely has a continuous family of scale invariant theories.

The main goal of this paper is to exploit the MERA tensor network and provide a systematic numerical approach to solve the corresponding critical data along the quantum phase transition in the $\mathbb{Z}_3$ chiral clock model. As we show, MERA can provide much more information about the scaling data than existing numerical approaches. However, while we hoped to be able to verify the non-trivial slow flow, we instead find that the scaling dimensions become unreliable and we cannot definitely establish whether there is another relevant operator for general $\theta$ describing the slow flow.

\subsection{Phase diagram}
\label{sec:phase-diagram}

The $\mathbb{Z}_3$ chiral clock model exhibits a rich phase structure in the three-parameter space $(f,\theta,\phi)$. Previous studies have identified three distinct many-body phases\cite{Zhuang:2015fpa,Samajdar:2018xkm}: (i) a \textbf{topological phase} that is gapped and threefold-degenerate, supporting $\mathbb{Z}_3$ parafermionic zero modes at the boundaries; (ii) a \textbf{trivial (paramagnetic) phase} that is gapped and non-degenerate; and (iii) an \textbf{incommensurate (IC) critical phase} that forms a gapless Luttinger liquid with central charge $c=1$. 

The onsite $\mathbb{Z}_3$ symmetry of the model is generated by
\[ Q = \prod_j \tau_j, \ \ \ \ Q \sigma_j Q^\dagger = \omega \sigma_j, \ \ \ \ \omega = e^{2\pi i /3}, \]
which cyclically permutes the three clock states on each site. Only the ordered phase (topological in the parafermion description) exhibits spontaneous breaking of the onsite \(\mathbb{Z}_3\) symmetry generated by \(Q=\prod_j\tau_j\), \emph{together with} one-site lattice translations \(T\) (reduced to the period-3 subgroup \(T^3\)). In spin variables this appears as long-range three-sublattice order, e.g. \(\langle\sigma_j\rangle = m\,\omega^{j+j_0}\) with \(\omega=e^{2\pi i/3}\) and \(m\neq 0\) in the thermodynamic limit. By contrast, the trivial paramagnet and the incommensurate Luttinger liquid preserve \(\mathbb{Z}_3\) and lattice translations; the latter displays algebraic correlations at an incommensurate wavevector rather than true long-range order~\cite{Zhuang:2015fpa,Samajdar:2018xkm,Whitsitt:2018hbw}.

In the full parameter space, for weak chirality ($\theta,\phi\lesssim\pi/6$), the topological and trivial phases meet directly along a Potts-type critical sheet with central charge $c=4/5$. Beyond this regime, the IC liquid intervenes, bounded by Kosterlitz–Thouless (KT) and Pokrovskii–Talapov (PT) transition surfaces~\cite{Zhuang:2015fpa}.

\begin{figure}[htbp]
    \centering
    
    \begin{subfigure}{0.5\textwidth}
        \includegraphics[width=0.8\textwidth]{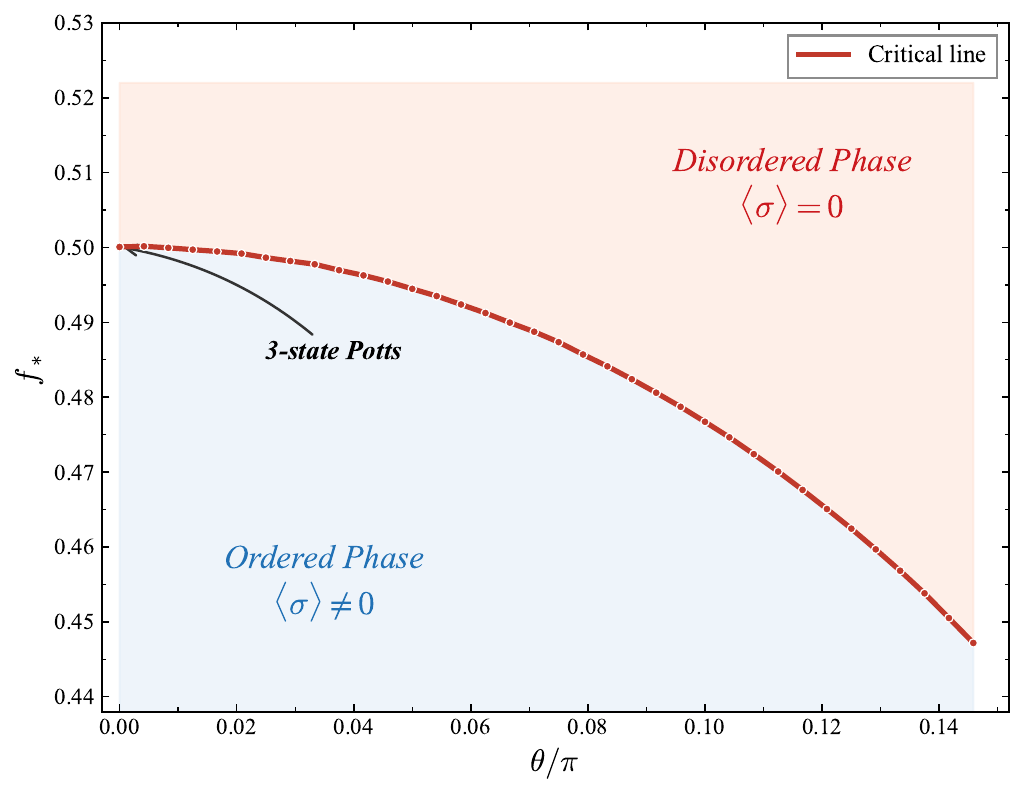}
        \caption{Phase diagram of $\mathbb{Z}_3$ chiral clock model}
    \end{subfigure}
    
    \begin{subfigure}{0.45\textwidth}
        \includegraphics[width=\textwidth]{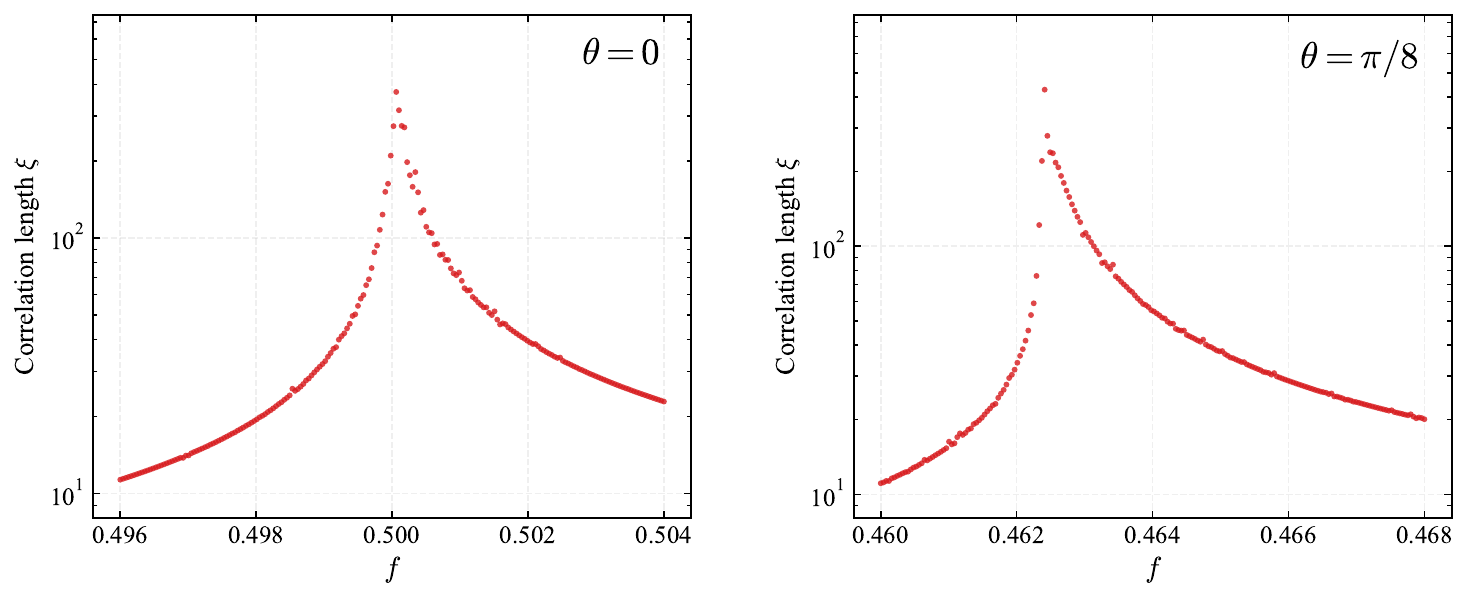}
        \caption{Correlation length $\xi$ vs. $f$ for $\theta = 0$ and $\theta = \pi/8$}
    \end{subfigure}%
    
    \caption{Phase diagram and correlation length analysis of the $\mathbb{Z}_3$ chiral clock model. (a) Critical line $f_c(\theta)$ in the weak chirality regime at $\phi = 0$, showing the direct transition between topological and trivial phases. (b) Correlation length $\xi$ as a function of coupling $f$ for representative values of the chiral parameter, computed using infinite DMRG (iDMRG) with bond dimension $\chi = 100$. The divergence of the correlation length identifies the critical point location $f_c$.}
    \label{fig:phase-diagram}
\end{figure}

In this work, we focus on the weak chirality regime with $\phi = 0$ and $\theta \in [0, \pi/8]$, where the system exhibits a direct continuous quantum phase transition between the topological and trivial phases. Figure \ref{fig:phase-diagram}(a) shows the phase diagram in the $(f,\theta)$ plane within this parameter range. The critical line $f_c(\theta)$ separates the topological phase (small $f$) from the trivial paramagnetic phase (large $f$). At $\theta=0$ and $f=0.5$, the system corresponds to the critical 3-state Potts model with known conformal field theory description. As $\theta$ increases, the critical coupling varies smoothly, and the system develops anisotropic scaling behavior that breaks conformal invariance.

\paragraph{Critical point determination via correlation length analysis.}
The critical points are determined using infinite DMRG (iDMRG) by systematically computing the correlation length $\xi(f)$ and identifying its divergence. Figure \ref{fig:phase-diagram}(b) demonstrates this procedure for two representative cases: $\theta = 0$ (conformal Potts point) and $\theta = \pi/8$ (non-conformal regime). The correlation length exhibits a clear peak at the critical coupling $f_c$, enabling precise determination of the phase boundary. This systematic approach provides the foundation for our subsequent MERA analysis of scaling properties across the conformal-to-non-conformal transition.

\paragraph{Topological nature and parafermions.}
At small $f$ the model enters a gapped phase which, in the clock-spin language, is naturally viewed as an ordered phase. However, it also admits a complementary ``topological'' interpretation in terms of boundary physics after a nonlocal Jordan--Wigner-type mapping to a $\mathbb{Z}_3$ parafermion chain~\cite{Fendley:2012vv}. In the parafermion representation, this phase corresponds to a gapped bulk together with exponentially localized (near-)zero-energy boundary modes that are $\mathbb{Z}_3$ parafermions---generalizations of Majorana zero modes which, in suitable architectures, can support non-Abelian braiding operations~\cite{Alicea:2015hja}. The existence of these parafermion edge modes is protected as long as the discrete $\mathbb{Z}_3$ symmetry is preserved and the bulk gap remains open. Consequently, an open chain exhibits a low-energy three-state manifold whose splitting is exponentially small in system size (and becomes exactly threefold degenerate in the thermodynamic limit)~\cite{Zhuang:2015fpa}. This boundary-origin quasi-degeneracy is distinct from conventional symmetry-breaking degeneracy: it is tied to edge modes and is therefore sensitive to boundary conditions (e.g., it need not appear for periodic chains), whereas symmetry-breaking degeneracy is a bulk property associated with a local order parameter. This protection and boundary control make parafermion zero modes of particular interest for topological quantum computation applications.

The parameter regime choice enables a systematic MERA analysis of the scaling-invariant critical theory while avoiding the computational complexities associated with incommensurate phases at larger chirality angles. This provides an ideal setting for studying the evolution from conformal ($\theta = 0$) to non-conformal ($\theta \neq 0$) critical behavior.

\subsection{Anisotropic scale-invariant field theory}
For systems with broken Lorentz invariance, the conventional conformal field theory framework breaks down, necessitating a more general theoretical description~\cite{Baiguera:2023wjf,Fradkin:2013ufw}. For an anisotropic (non-relativistic) scaling-invariant system \cite{Lifshitz:1941tl}, time and space do not scale in the same way. The symmetry algebra in these theories is significantly reduced compared to conformal field theories: while CFTs possess the full conformal group including translations, rotations, scale transformations, and special conformal transformations, non-relativistic theories retain only a subset consisting of time translations and spatial translations, spatial rotations, and anisotropic scaling transformations. Crucially, the Lorentz boost symmetry and special conformal transformations are absent, which fundamentally alters the structure of correlation functions and operator algebras. The loss of boost invariance means that the lightcone structure familiar from relativistic theories is replaced by a more general causal structure determined by the dynamical exponent $z$. In this kind of system, it is natural to ask if we can have boost-like symmetry generators and analogue of OPEs, though the answers are generally more complex than in the conformal case. 

The scale transformations of an anisotropic scale-invariant theory are $t \to \lambda^z t,  x \to \lambda x$. Here $z$ is the dynamical critical exponent, which describes the asymmetries in time and space~\cite{Hohenberg:1977ym}. The symmetry generators of this system include $P_i = -i\partial_i, H = -i \partial_t,J_{ij} = -i(x_i \partial_j - x_j \partial_i), D = -i(zt\partial_t + x^i \partial_i)$. The operator scaling behavior is $O(\lambda x, \lambda^z t) = \lambda^{-\Delta} O(x, t)$, where $\Delta$ is the scaling dimension. 

For systems with non-unity dynamical critical exponents, we can define an analogue of the OPE formula. A scaling operator transforms as $\phi(t,x) \to \lambda^{-\Delta} \phi( \lambda^z t, \lambda x)$. The operator spectrum in non-conformal theories exhibits a more complex structure than in CFTs, with scaling dimensions that depend on both the spatial and temporal derivative content~\cite{Baiguera:2023wjf}. In this context, the notion of "primary" operators is generalized: rather than being defined by their transformation properties under the full conformal group (which is absent), they are now characterized as operators that are not descendants of other operators under the reduced symmetry algebra. For such a scaling operator $\Phi$ with dimension $\Delta$, 

\begin{itemize}
  \item Time descendants $(\partial_t)^n \Phi$ have dimension $\Delta + nz$
  \item Space descendants $(\partial_i)^m \Phi$ have dimension $\Delta + m$
  \item Mixed descendants $(\partial_t)^n (\partial_i)^m \Phi$ have dimension $\Delta + nz + m$
\end{itemize}

This non-uniform structure forms an anisotropic tower of descendants (spatial derivatives raise $\Delta$ by 1 while temporal derivatives raise it by $z$), reflecting the underlying $z \neq 1$ scaling.

The two-point correlation function in non-conformal theories provides the most fundamental characterization of critical behavior. Unlike in CFT where scaling symmetry fully determines the functional form, anisotropic scaling leads to correlation functions that depend on universal scaling functions:
\begin{equation}
\langle O_i(x_1,t_1) O_j(x_2,t_2)\rangle
= |x_{12}|^{-\Delta_i-\Delta_j}\,
\mathcal{F}_{ij}\!\left(\frac{t_{12}}{|x_{12}|^z}\right),
\end{equation}
where $x_{12}=x_1-x_2$, $t_{12}=t_1-t_2$, and $\mathcal{F}_{ij}$ is a universal scaling function that encodes the anisotropic spacetime structure~\cite{Baiguera:2023wjf}. The scaling function $\mathcal{F}_{ij}$ cannot be determined purely from symmetry considerations and must be computed or measured.

The extension of the operator product expansion (OPE) to scale-invariant but non-conformal theories remains an active area of research with limited established results. For a pair of scalar scaling operators $O_a$ and $O_b$, translation invariance and anisotropic scaling allow the short-distance expansion
\begin{equation}
\begin{aligned}
O_a(t,x)\,O_b(0,0)
&\sim
\sum_{c}
|x|^{\Delta_c-\Delta_a-\Delta_b}\,
f_{ab}^{\;c}\!\left(\frac{t}{|x|^{z}}\right)
\\
&\qquad\times O_c(0,0)
+\dots,
\end{aligned}
\label{eq:anisotropic_OPE}
\end{equation}
where $f_{ab}^{\;c}(u)$ are dimensionless scaling functions of the invariant ratio $u=t/|x|^{z}$, and the dots denote descendant contributions involving additional powers of $x$ and derivatives. The leading channel $c=\mathbf{1}$ corresponds to the identity operator and reproduces the two-point scaling function, $\mathcal{F}_{ab}(u)=f_{ab}^{\;\mathbf{1}}(u)$. In a relativistic CFT with $z=1$, conformal symmetry forces $f_{ab}^{\;c}(u)$ to be independent of $u$, reducing Eq.~(\ref{eq:anisotropic_OPE}) to the familiar primary OPE with constant structure constants $C_{ab}^{\;c}$. For general anisotropic scaling with $z\neq 1$, however, the functions $f_{ab}^{\;c}(u)$ are not fixed by symmetry and are poorly understood analytically. This lack of a complete OPE framework for non-relativistic theories represents a significant gap in our analytical understanding and motivates the development of systematic numerical approaches to extract generalized OPE data directly from correlation functions.

In the following we will restrict to equal-time insertions, $t=0$, and define generalized equal-time OPE amplitudes
\begin{equation}
C_{ab}^{\;c} \equiv f_{ab}^{\;c}(0),
\end{equation}
so that the leading equal-time OPE takes the CFT-like power-law form
\begin{equation}
O_a(0,x)\,O_b(0,0)
\sim
\sum_{c} C_{ab}^{\;c}\,
|x|^{\Delta_c-\Delta_a-\Delta_b}\,O_c(0,0).
\end{equation}
These coefficients $C_{ab}^{\;c}$ are the generalized OPE data that we will extract from MERA in the next section.

\subsection{Critical exponents}
Non-conformal quantum phase transitions are characterized by critical exponents that deviate from conformal field theory predictions~\cite{Dutta:2015qay,Sachdev:2011xva}. The key exponents are the dynamical critical exponent $z$, defined by the anisotropic scaling $t \to \lambda^z t, x \to \lambda x$, and the correlation length exponent $\nu$~\cite{Cardy:1996yz}. Near the critical point $f \to f_c$, these exponents govern the scaling behavior according to
\begin{equation}
    \Delta E \sim |f-f_c|^{z\nu}, \quad \xi \sim |f-f_c|^{-\nu}
\end{equation}
where $\Delta E$ is the energy gap and $\xi$ is the correlation length.

Previous studies have employed finite-size scaling analysis to extract critical exponents from numerical simulations~\cite{Affleck:1987cy}. High-precision DMRG calculations determine $z$ and $\nu$ by analyzing the system-size dependence of energy gaps and correlation lengths near criticality~\cite{Samajdar:2018xkm}.

The MERA tensor network provides an alternative systematic approach to extract these critical exponents directly from the operator spectrum. The dynamical critical exponent can be determined from the scaling dimensions of temporal and spatial descendants of primary operators: $z = \Delta_{\partial_t \Phi} - \Delta_{\partial_x \Phi} + 1$. The correlation length exponent follows from the hyperscaling relation $\frac{1}{\nu} = 1 + z - \Delta_\epsilon$, where $\Delta_\epsilon$ is the scaling dimension of the relevant driving operator. To derive this, note that the tuning parameter $g = f - f_c$ transforms under the RG as $g \to \lambda^{y_g} g$ with $y_g = d + z - \Delta_\epsilon$, where $d$ is the spatial dimension and $d+z$ is the effective spacetime dimension. Since the correlation length diverges as $\xi \sim |g|^{-\nu}$ and $\xi$ is a purely spatial quantity ($\xi \to \lambda\,\xi$ under rescaling, not $\lambda^z\xi$), matching gives $\nu = 1/y_g = 1/(d + z - \Delta_\epsilon)$. Setting $d=1$ yields $1/\nu = 1 + z - \Delta_\epsilon$~\cite{Sachdev:2011xva}.

\section{Multiscale Entanglement Renormalization Ansatz}
\label{sec:methods}

The Multiscale Entanglement Renormalization Ansatz (MERA) is a tensor network ansatz specifically designed to efficiently represent the ground states of quantum many-body systems with scale-invariant properties~\cite{Vidal:2007sv,Evenbly:2009jq,Pfeifer:2009fg}. MERA has proven particularly effective for studying quantum critical systems due to its ability to capture the logarithmic violations of the area law for entanglement entropy that characterize critical ground states~\cite{Calabrese:2004eu,Ryu:2006bv}. The key insight underlying MERA is that quantum critical systems possess hierarchical entanglement structures that can be systematically disentangled through a multiscale renormalization procedure~\cite{Vidal:2008zz}.

\subsection{Tensor Network Architecture}
\label{sec:mera-architecture}

MERA is a layered tensor network built from two types of tensors—\emph{disentanglers} and \emph{isometries}—arranged in an alternating, tree-like hierarchy~\cite{Orus:2014tka,Cirac:2021kly}. Disentanglers remove short-range entanglement between neighboring sites, while isometries perform a coarse-graining that reduces the number of degrees of freedom by a factor of two at each layer. This hierarchical layout naturally captures scale-invariant features and long-range correlations characteristic of critical systems.

\begin{itemize}
\item \textbf{Disentanglers} ($u$ tensors): Four-index unitary tensors that locally remove entanglement between adjacent degrees of freedom without changing the total number of degrees of freedom.

\item \textbf{Isometries} ($\omega$ and $v$ tensors): Three-index tensors satisfying $\omega^\dagger\omega=v^\dagger v=\mathbb{I}$ that implement a coarse-graining transformation, reducing the number of effective degrees of freedom by a factor of two per layer.
\end{itemize}

In this work we employ a \emph{modified binary} MERA. Each layer still effects a $2\!\to\!1$ coarse-graining, but we alternate two species of isometries, which we denote $\omega$ and $v$, along the chain. This alternation preserves translation invariance with period two and accommodates the absence of reflection symmetry (chirality). The resulting causal cones have width two, which keeps contraction cost controlled while faithfully representing power-law correlations.

\begin{figure}[ht]
  \centering
  \includegraphics[width=0.42\textwidth]{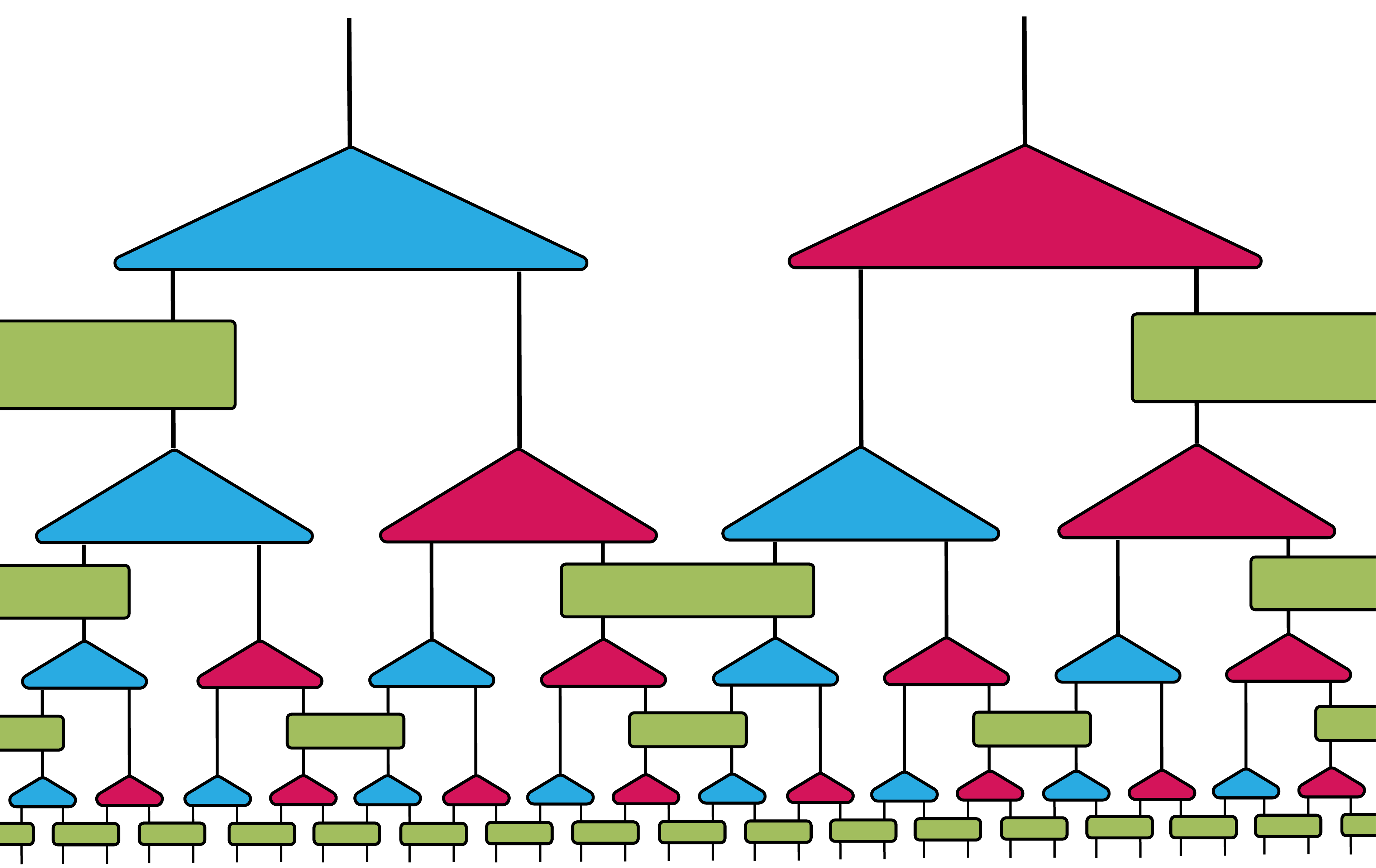}
  \caption{Schematic of the two-site modified binary MERA.  Blue and red triangles represent the two isometry species \(\omega\) and \(v\); green rectangles are disentanglers \(u\).  Each layer realises a \(2\!\to\!1\) coarse-graining map, and removing every second disentangler reduces the causal width from three sites to two.}
  \label{fig:modified-binary-mera}
\end{figure}

We only impose the unitarity/isometry constraints on $u$ and the two isometries $\omega, v$. The per-layer scale factor is $b=2$, which fixes the base of all logarithms in the definition of scaling dimensions. For clarity of notation used later: we reserve $\mu_\alpha$ for eigenvalues of scaling superoperators (not tensor names); disentanglers are $u$ and isometries are $\omega, v$.

Viewed from bottom to top, the MERA implements a real-space renormalization group (RG) transformation for quantum many-body systems~\cite{Cardy:1996yz,Sachdev:2011xva}. Each layer corresponds to an RG step in which short-range entanglement is first removed by disentanglers and then the lattice is coarse-grained by isometries. Iterating this procedure systematically eliminates high-energy degrees of freedom while preserving the long-range correlations that encode critical behavior~\cite{Haegeman:2011uy}. At a scale-invariant fixed point (achieved variationally in our optimization), the repeating tensor pattern underpins the extraction of universal scaling data; the precise mechanism is developed in the next subsection.

\paragraph*{Computational trade-offs.}
Relative to the original binary and ternary MERA networks, the modified binary MERA layout offers three practical advantages: it lowers the leading contraction cost for local observables to $\mathcal{O}(\chi^{7})$ (versus $\mathcal{O}(\chi^{9})$ for binary and $\mathcal{O}(\chi^{8})$ for ternary); it achieves a favorable cost–accuracy balance, with benchmarks indicating reduced variational energy errors at comparable floating-point budget once $\chi$ is moderately large; and it yields a more compact causal structure by removing every second disentangler, reducing the causal width from three sites to two, thereby shortening ascending/descending superoperator networks and decreasing the per-layer memory footprint~\cite{Evenbly:2011dki}.

\subsection{Extracting Scaling Data from MERA}
\label{sec:scaling-data}

We extract universal scaling data within the modified binary architecture introduced above. Because each layer halves the number of sites, the fundamental scale factor is \(b=2\); causal cones have width two.

\paragraph*{Why MERA extracts scaling data.} MERA implements a real-space renormalization group (RG) using local disentanglers (unitaries) to remove short-range entanglement and isometries to coarse-grain. At a quantum critical point, the optimized tensors flow to a scale-invariant fixed point so that the same circuit (up to a period-two spatial unit cell here) recurs at every scale. This fixed point induces a linear RG map on local operators (the ascending superoperator). A scaling operator \(\phi_\alpha\) is an eigenoperator of this map and rescales by a factor related to its scaling dimension. Long-distance correlators reduce, after \(\ell=\log_2|x|\) layers, to short-distance contractions inside a bounded causal cone, reproducing power laws \(\langle\phi_\alpha(x)\phi_\alpha(0)\rangle\propto |x|^{-2\Delta_\alpha}\). Three-point contractions at a common layer similarly encode OPE coefficients. Relative to MPS-based methods, MERA natively represents the logarithmic violation of the area law at criticality while keeping contraction cost polylogarithmic in distance.

\paragraph*{Scaling superoperators and scaling dimensions.} In a modified binary MERA, a one-layer ascent toggles sublattices. Let \(\mathcal{S}_A\!: A\to B\) and \(\mathcal{S}_B\!: B\to A\). The physical scaling spectrum is read from the two-layer composites
\begin{equation}
  \mathcal{S}^{(2)}_{A} \equiv \mathcal{S}_{B}\!\circ\!\mathcal{S}_{A},\qquad
  \mathcal{S}^{(2)}_{B} \equiv \mathcal{S}_{A}\!\circ\!\mathcal{S}_{B},
\end{equation}
which map each sublattice to itself with overall scale factor \(b^2\) (\(b=2\) per layer). Diagonalizing \(\mathcal{S}^{(2)}_{A}\) (without imposing symmetry blocks) yields
\begin{equation}
  \mathcal{S}^{(2)}_{A}[\phi^{A}_{\alpha}] = \mu_{\alpha}\,\phi^{A}_{\alpha},
  \qquad \Delta_{\alpha} = -\tfrac{1}{2}\,\log_{2}\mu_{\alpha} 
  = -\frac{1}{2}\,\frac{\ln \mu_{\alpha}}{\ln 2}.
\end{equation}
Given \(\phi^{A}_{\alpha}\), the partner on sublattice B is reconstructed by one ascent
\begin{equation}
  \phi^{B}_{\alpha} = \frac{\mathcal{S}_{A}[\phi^{A}_{\alpha}]}{\sqrt{\mu_{\alpha}}},
\end{equation}
so that \(\mathcal{S}_{A}[\phi^{A}_{\alpha}] = \sqrt{\mu_{\alpha}}\,\phi^{B}_{\alpha}\) and \(\mathcal{S}_{B}[\phi^{B}_{\alpha}] = \sqrt{\mu_{\alpha}}\,\phi^{A}_{\alpha}\). This fixes a symmetric normalization for the one-step maps: each single ascent carries a factor \(\sqrt{\mu_{\alpha}}\) while toggling \(A\leftrightarrow B\). In other normalizations the single-step rescaling factors can differ (say \(\lambda^{(A)}_{\alpha}\) and \(\lambda^{(B)}_{\alpha}\)), but their product is fixed by the two-layer eigenvalue, \(\mu_{\alpha}=\lambda^{(A)}_{\alpha}\lambda^{(B)}_{\alpha}\). Because these single-step factors depend on normalization, we do not use them to report \(\Delta\); they serve only as optional diagnostics.

\begin{figure}[ht]
  \centering
  \begin{subfigure}{0.45\textwidth}
    \includegraphics[width=\textwidth]{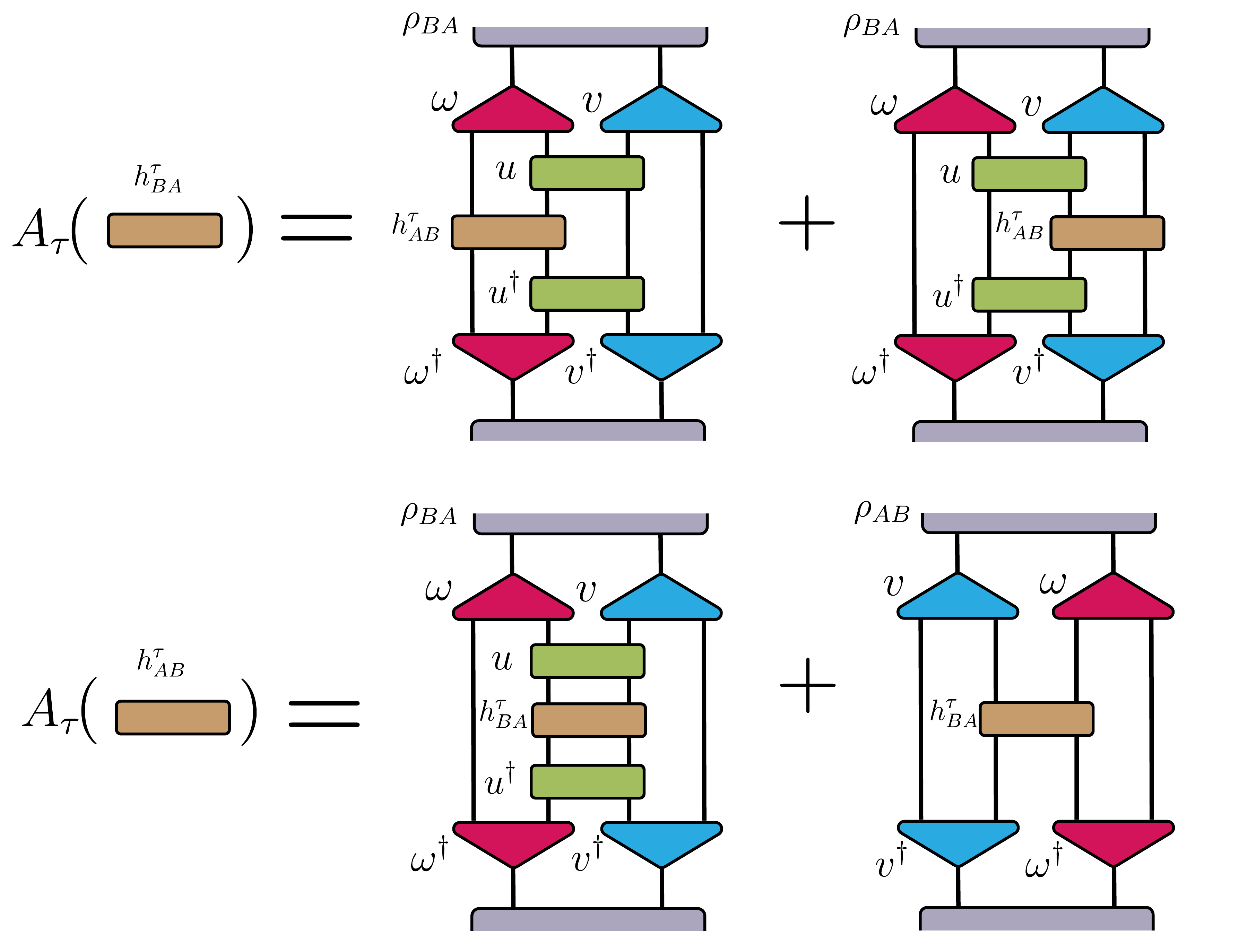}
    \caption{Modified binary ascending superoperator}
    \label{fig:modified-ascending}
  \end{subfigure}
  \begin{subfigure}{0.45\textwidth}
    \includegraphics[width=\textwidth]{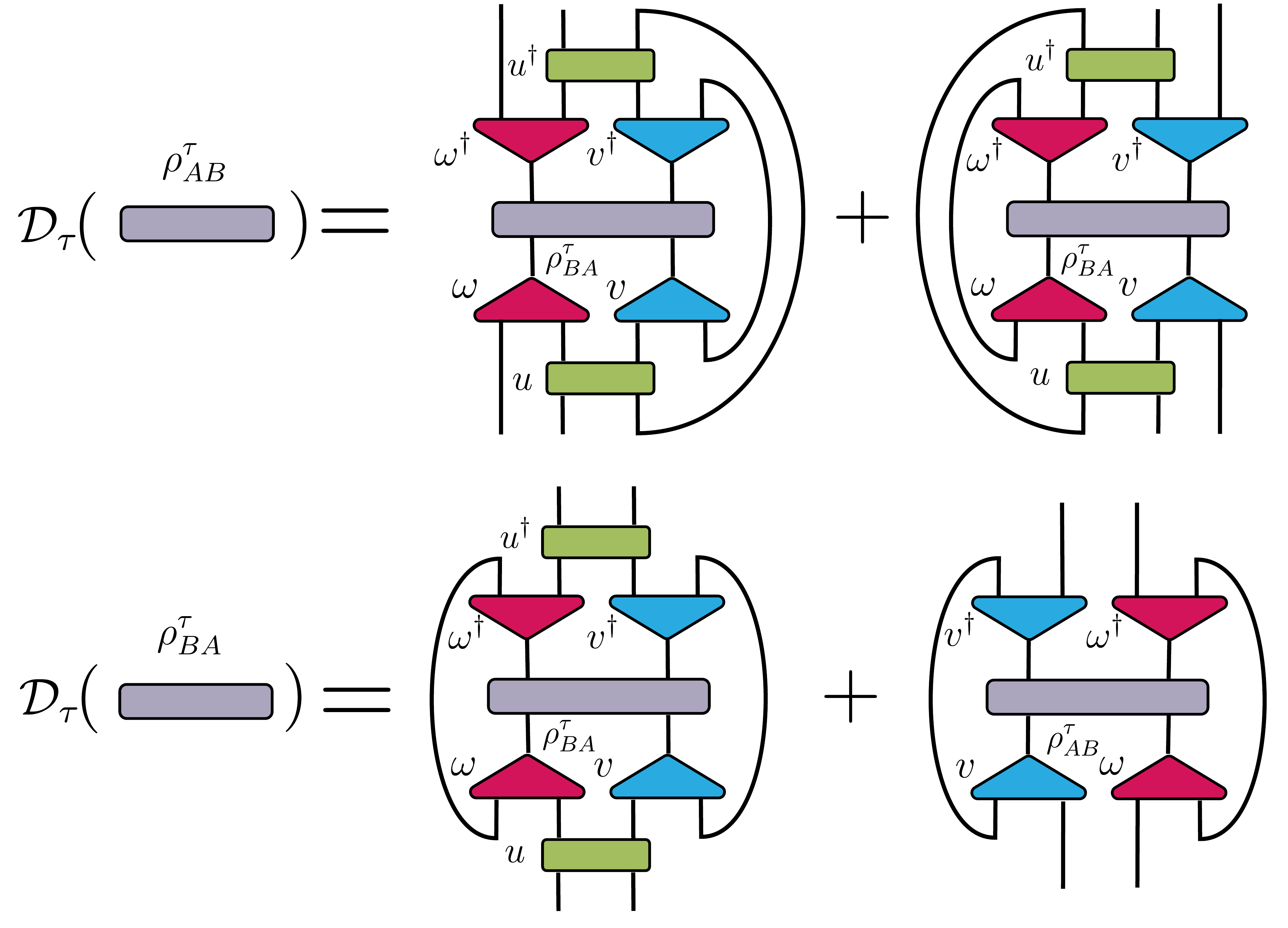}
    \caption{Modified binary descending superoperator}
    \label{fig:modified-descending}
  \end{subfigure}
  \caption{Modified binary MERA superoperators. (a) The ascending superoperator implements the scale transformation for operators. (b) The descending superoperator implements the inverse transformation for reduced density matrices.}
  \label{fig:modified-superoperators}
\end{figure}

\paragraph*{Operator normalization and identification.} We fix two-point normalization via the descending channel. With this convention \(C_{\alpha\beta 1}=\delta_{\alpha\beta}\), and derivative descendants obey the exact relation
\begin{equation}
  \Delta_{\partial_x \Phi} - \Delta_{\Phi} = 1,
\end{equation}
which provides an internal consistency check alongside conjugate-pair degeneracies.

\paragraph*{Critical exponents from scaling data.} The scaling spectrum also yields critical exponents. Using the exact spatial-derivative relation above, the dynamical exponent can be evaluated in two equivalent forms,
\begin{equation}
  z = \Delta_{\partial_t \Phi} - \Delta_{\partial_x \Phi} + 1 = \Delta_{\partial_t \Phi} - \Delta_{\Phi},
\end{equation}
and the correlation-length exponent follows the hyperscaling relation
\begin{equation}
  \frac{1}{\nu} = 1 + z - \Delta_{\epsilon},
\end{equation}
where $\epsilon$ is the relevant energy-density operator. In practice we read off $\Delta_{\Phi}$, $\Delta_{\partial_x \Phi}$, $\Delta_{\partial_t \Phi}$, and $\Delta_{\epsilon}$ from the eigenvalues of the two-layer ascending superoperator.

\begin{figure}[ht]
  \centering
  \includegraphics[width=0.25\textwidth]{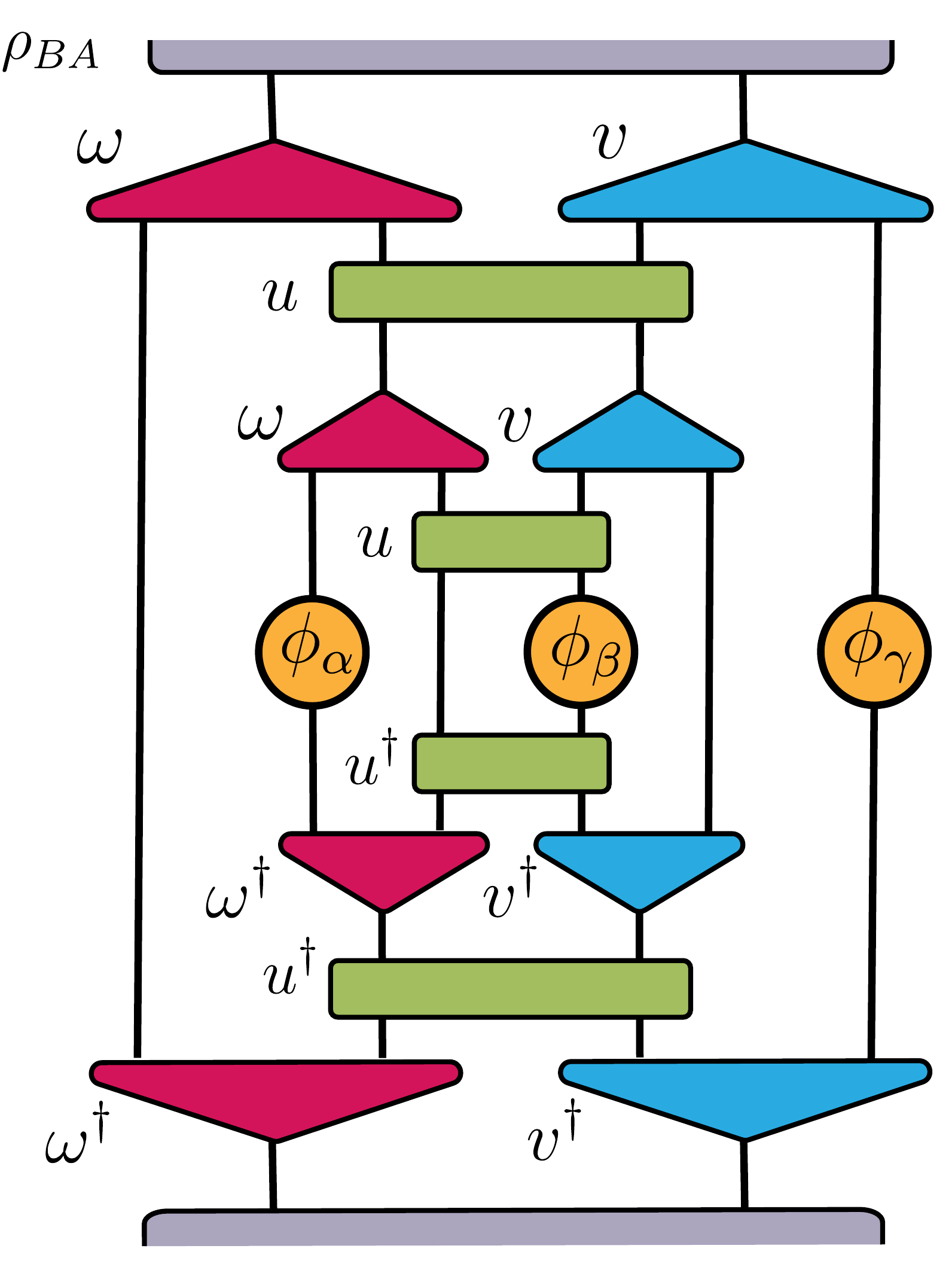}
  \caption{Three-point function contraction in the modified binary MERA. Inserting the scaling operators at a common layer confines the computation to their joint causal cone ($b=2$); contracting to the first common ancestor yields $\langle \phi_\alpha(x_1)\phi_\beta(x_2)\phi_\gamma(x_3)\rangle$, from which the OPE coefficient $C_{\alpha\beta\gamma}$ is extracted.}
  \label{fig:3pt_mera}
\end{figure}

\paragraph*{OPE coefficients from three-point contractions.} The MERA framework enables the extraction of the generalized equal-time OPE coefficients introduced above from three-point correlation functions. We consider equal-time correlators of scaling operators $O_a$, $O_b$ and $O_c$ and assume a CFT-inspired spatial dependence
\begin{equation}
\label{eq:3pt_MERA}
\langle O_a(x_1)\,O_b(x_2)\,O_c(x_3)\rangle
=
\frac{C_{abc}}
{|x_{12}|^{\Omega_{ab}}\,
 |x_{23}|^{\Omega_{bc}}\,
 |x_{13}|^{\Omega_{ac}}}\,,
\end{equation}
where $x_{ij}=x_i-x_j$ and $\Omega_{ab}=\Delta_a+\Delta_b-\Delta_c$, $\Omega_{bc}=\Delta_b+\Delta_c-\Delta_a$, $\Omega_{ac}=\Delta_a+\Delta_c-\Delta_b$.
By restricting to equal-time insertions, $t_1=t_2=t_3$, we eliminate any
dependence on the anisotropic scaling variable $t/|x|^z$ and can interpret
the amplitudes $C_{abc}$ as the equal-time OPE data
$C_{ab}^{\;c}=f_{ab}^{\;c}(0)$ defined in \eqref{eq:anisotropic_OPE}.
In a basis where the two-point functions of scaling operators are
orthonormal,
$\langle O_a(x)\,O_b(0)\rangle \propto \delta_{ab}\,|x|^{-2\Delta_a}$,
the three-point coefficients $C_{abc}$ coincide with the equal-time OPE
coefficients $C_{ab}^{\;c}$ appearing in the leading OPE
\begin{equation}
O_a(0,x)\,O_b(0,0)
\sim
\sum_{c} C_{ab}^{\;c}\,
|x|^{\Delta_c-\Delta_a-\Delta_b}\,O_c(0,0).
\end{equation}
Operationally, in the MERA implementation we insert the three operators
on a common layer of the network so that their support is contained in a
single joint causal cone, and then contract the tensors up to their first
common ancestor to evaluate the three-point correlator. For practical
implementation we use the contraction pattern shown in
Fig.~\ref{fig:3pt_mera}.

\subsection{Optimization Algorithm}
\label{sec:optimization}

The MERA tensors are optimized using a variational approach that minimizes the energy expectation value of the Hamiltonian~\cite{Evenbly:2009jq}. The optimization procedure alternates between updating the disentanglers and isometries, with each update involving the solution of a generalized eigenvalue problem. This iterative process continues until convergence is achieved, typically requiring several hundred to thousands of iterations depending on the desired precision~\cite{Evenbly:2009jq}.

The optimization process continues until convergence, typically measured by the stabilization of the ground state energy and the scaling dimensions extracted from the scaling superoperator (see \secref{subsec:mera_details}).

\section{Numerical Results}
\label{sec:results}

We present a systematic MERA analysis of the $\mathbb{Z}_3$ chiral clock model, investigating its critical properties across the phase diagram. Our numerical investigation yields four key results: (1) detailed operator spectrum revealing scaling dimensions of primary and descendant operators, (2) precise critical exponents including the dynamical critical exponent $z$ and correlation length exponent $\nu$, (3) operator product expansion coefficients characterizing the fusion algebra, and (4) the effective central charge $c_{\mathrm{eff}}$ characterizing entanglement degrees of freedom along the critical line.

\subsection{Phase Diagram and Critical Points}
\label{sec:phase-diagram-results}

To map quantum phase transitions in the $\mathbb{Z}_3$ chiral clock model, we employed the infinite Density Matrix Renormalization Group (iDMRG) method across the $(\theta,f)$ parameter space. Critical points are identified by locating the values of $f$ where the correlation length $\xi$ reaches its maximum for each fixed $\theta$ parameter.

The critical coupling $f_c(\theta)$ exhibits non-monotonic dependence on the chiral parameter $\theta$. We highlight several key observations. At $\theta = 0$, we recover the known critical point $f_c = 0.5$ for the 3-state Potts model, validating our approach. As $\theta$ increases, the critical coupling reflects the influence of chiral interactions on the phase boundary. The direct Potts-type critical line persists up to $\theta \approx \pi/6$; for larger chirality, the transition splits and an incommensurate (floating) critical Luttinger-liquid phase ($c=1$) intervenes between the topological (ordered) and trivial (paramagnetic) phases. The disordered--IC boundary is of Kosterlitz--Thouless type, while the ordered--IC boundary is of Pokrovskii--Talapov type~\cite{Zhuang:2015fpa,Samajdar:2018xkm,Whitsitt:2018hbw}. 

The iDMRG method provides precise determination of critical couplings by systematically scanning the correlation length as a function of $f$, where the divergence of $\xi$ signals the quantum critical point.

\subsection{Conformal Benchmark: 3-State Potts Model ($\theta = 0$)}
\label{sec:potts-benchmark}

We validate our MERA methodology by reproducing known results of the critical 3-state Potts model at $\theta = 0$. This conformal field theory benchmark provides essential verification before examining non-conformal effects.

At the critical point $f_c = 0.5$, the 3-state Potts model exhibits conformal invariance with central charge $c = 4/5$. Our MERA calculations successfully reproduce theoretical predictions. We use the lattice–CFT operator dictionary in \secref{sec:models} to identify the fields $\sigma$ ($\Delta=2/15$), $\epsilon$ ($\Delta=4/5$), and $\psi$ ($\Delta=4/3$).

\begin{table}[htb]
\centering
\small
\begin{tabular}{lccc}
\hline
Field & Theory & MERA & Error \\
\hline
$\sigma, \sigma^\dagger$ & $2/15$ & $0.1348$ & $1.11\%$ \\
$\epsilon$ & $4/5$ & $0.8221$ & $2.76\%$ \\
$\partial_x \sigma$, $\partial_x \sigma^\dagger$ & 17/15 & $1.1391$ & $0.51\%$ \\
$\partial_t \sigma$, $\partial_t \sigma^\dagger$ & 17/15 & $1.1444$ & $0.98\%$ \\
$\psi, \psi^\dagger$ & $4/3$ & $1.3288$ & $0.34\%$ \\
\hline
\end{tabular}
\caption{Low-dimension scaling operators for 3-state Potts model.}
\label{tab:potts-scaling-dimensions}
\end{table}

\begin{table}[htb]
\centering
\small
\begin{tabular}{lccc}
\hline
OPE Coefficient & Theory & MERA & Error \\
\hline
$C_{\sigma\sigma^{\dagger}\epsilon}$ & $0.5462$ & $0.5391$ & $1.30\%$ \\
$C_{\epsilon\epsilon\epsilon}$ & $0$ & $0.00086$ & -- \\
$C_{\sigma\sigma\sigma}$ & $1.0924$ & $1.0866$ & $0.53\%$ \\
$C_{\psi\psi^{\dagger}\epsilon}$ & -- & $0.0293$ & -- \\
$C_{\sigma^{\dagger}\psi\epsilon}$ & $0.6667$ & $0.6491$ & $2.64\%$ \\
$C_{\sigma\sigma\psi}$ & $0.3333$ & $0.3354$ & $0.63\%$ \\
$C_{\psi\psi\psi}$ & -- & $1.3342$ & -- \\
\hline
\end{tabular}
\caption{OPE coefficients for 3-state Potts model.}
\label{tab:potts-ope-coefficients}
\end{table}

\begin{table}[htb]
\centering
\small
\begin{tabular}{lccc}
\hline
Exponent & Theory & MERA & Error (\%) \\
\hline
$z$ & $1$ & $0.9975$ & $0.25$ \\
$1/\nu$ & $6/5$ & $1.1762$ & $1.98$ \\
$\beta$ & $1/9$ & $0.1104$ & $0.63$ \\
\hline
\end{tabular}
\caption{Critical exponents for 3-state Potts model. The order-parameter exponent $\beta$ follows from the hyperscaling relation $\beta = \nu\,\Delta_\sigma$.}
\label{tab:potts-critical-exponents}
\end{table}

The excellent agreement between MERA results and known theoretical values demonstrates the reliability of our tensor network approach for extracting universal scaling properties. As shown in Tables~\ref{tab:potts-scaling-dimensions}, \ref{tab:potts-ope-coefficients}, and \ref{tab:potts-critical-exponents}, the Potts-point scaling dimensions deviate from exact CFT values by up to $\sim 2.8\%$ (dominated by $\Delta_\epsilon$); this offset represents a deterministic finite-$\chi$ bias common to all $\theta$. The $\theta$-dependent statistical uncertainty, measured by internal-consistency residuals, is below $0.5\%$ for all scaling dimensions. Critical exponents are within $2\%$ of theoretical predictions.

\subsection{Chiral Clock Model ($\theta \neq 0$)}
\label{sec:chiral-effects}

Having validated our approach with the Potts benchmark, we examine chiral deformation effects. For $\theta \neq 0$, the system deviates from conformal invariance, exhibiting rich non-conformal scaling behavior with $z \neq 1$.

\subsubsection{Scaling Dimension Evolution}
\label{sec:scaling-dimensions}

The complete operator spectrum exhibits systematic variation with the chiral parameter, reflecting continuous deformation of the underlying critical theory. While we retain the conventional notation ($\sigma$, $\epsilon$, $\psi$) inherited from the Potts point for continuity, it is important to note that for $\theta \neq 0$, these labels represent the evolved operators that continuously deform from their Potts counterparts rather than maintaining their original physical interpretations as spin, energy, and parafermion fields. Operator identification for $\theta > 0$ is based on the joint use of three criteria: (i)~\textit{Continuity}---we scan from $\theta=0$ in small steps $\Delta\theta = \pi/48$, tracking the continuous evolution of each eigenvalue; (ii)~\textit{$\mathbb{Z}_3$ charge}---eigenvalues are classified by the quantum number of $Q = \prod_j \tau_j$ (the spin field $\sigma$ and its spatial derivative $\partial_x\sigma$ carry charge~1, the energy operator $\epsilon$ carries charge~0, and the parafermion $\psi$ carries charge~2); (iii)~\textit{Degeneracy pattern}---conjugate pairs such as $\sigma/\sigma^\dagger$ are required to be exactly degenerate, providing a further consistency check.

\begin{figure}[htb]
    \centering
    \includegraphics[width=0.45\textwidth]{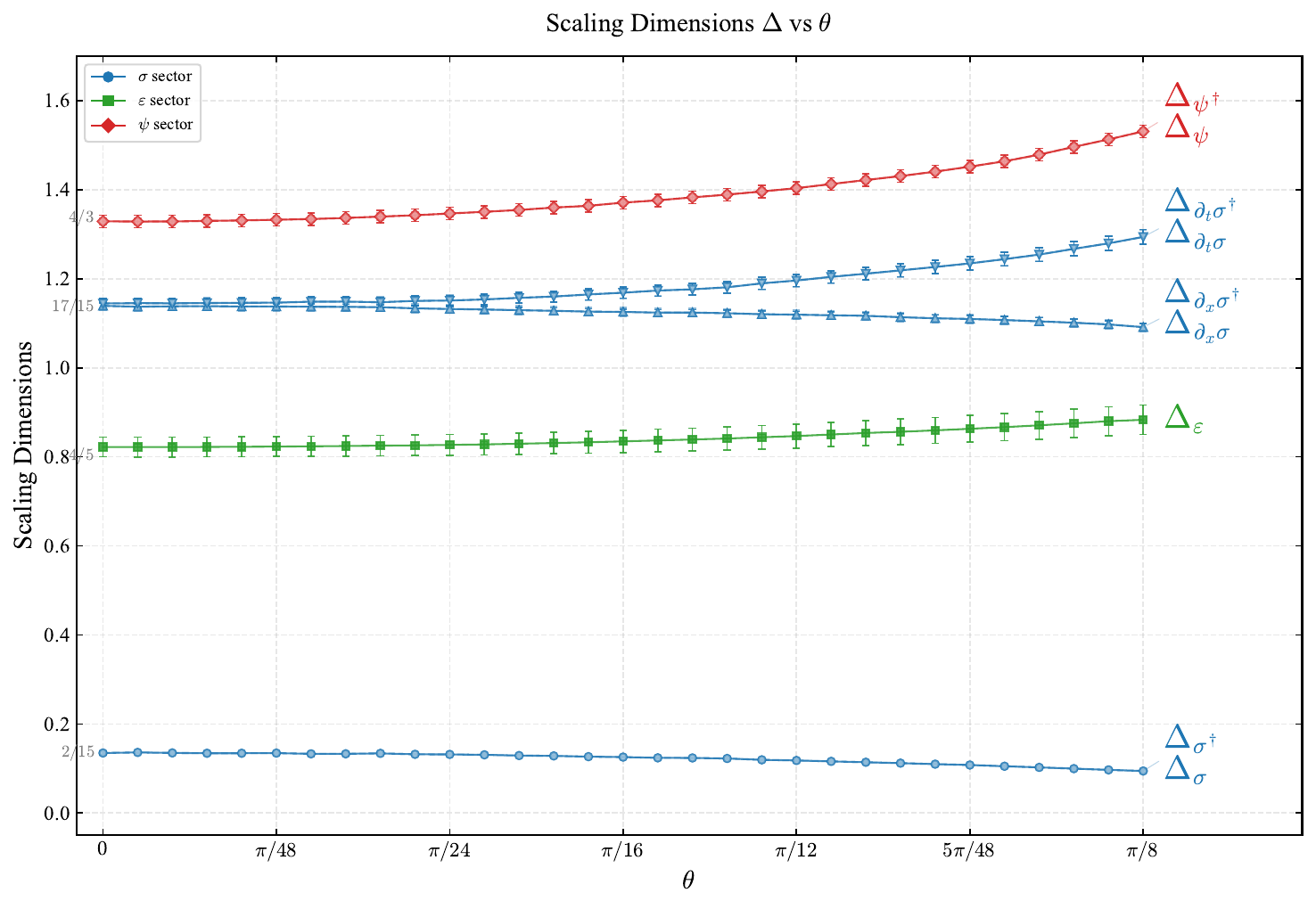}
    \caption{Evolution of scaling dimensions with chiral parameter $\theta$, grouped by physical sector ($\sigma$ in blue, $\epsilon$ in green, $\psi$ in red). Potts theoretical values ($2/15$, $4/5$, $17/15$, $4/3$) are indicated on the left axis. The splitting between temporal and spatial descendants of the spin field directly manifests the emergence of anisotropic scaling with $z \neq 1$. Error bars are determined by the comprehensive three-method error analysis detailed in Appendix~\ref{subsec:error_analysis}.}
    \label{fig:scaling-dimensions-evolution}
\end{figure}

The operator spectrum shown in Figure~\ref{fig:scaling-dimensions-evolution} comprises five non-trivial operator families grouped by physical sector: the spin field $\sigma$ and its conjugate $\sigma^\dagger$ (2-fold degenerate), energy density $\epsilon$, spatial descendant $\partial_x \sigma$ (2-fold degenerate), temporal descendant $\partial_t \sigma$ (2-fold degenerate), and parafermion field $\psi$ and its conjugate $\psi^\dagger$ (2-fold degenerate).

At $\theta = 0$, all scaling dimensions recover known theoretical values for the critical 3-state Potts model: $\Delta_\sigma = 2/15$, $\Delta_\epsilon = 4/5$, and $\Delta_\psi = 4/3$. As $\theta$ increases from 0 to $\pi/8$, the scaling dimensions evolve smoothly but in different directions: $\Delta_\sigma$ decreases monotonically (from $2/15$ to $\approx 0.094$), while $\Delta_\epsilon$ and $\Delta_\psi$ increase steadily above their Potts values. The growing separation between $\Delta_{\partial_t \sigma}$ and $\Delta_{\partial_x \sigma}$ demonstrates deviation of the dynamical critical exponent $z = \Delta_{\partial_t \sigma} - \Delta_\sigma$ from unity, signaling conformal symmetry breakdown. 

We observe that $\Delta_{\partial_x \sigma} - \Delta_{\sigma}$ deviates slightly above unity with increasing $\theta$. Since the exact relation $\Delta_{\partial_x \Phi} - \Delta_\Phi = 1$ holds for any value of $z$ (it follows purely from spatial rescaling $\partial_x \to \lambda^{-1}\partial_x$), this deviation is not a physical effect but rather quantifies the systematic error introduced by the finite bond dimension $\chi=36$ in the non-conformal regime. We use the magnitude of this deviation as one of our internal consistency diagnostics (Method~2 of Appendix~\ref{subsec:error_analysis}), and it is incorporated into the error bars reported throughout.

\subsubsection{Critical Exponents and Dynamical Scaling}
\label{sec:critical-exponents}

Critical exponents, particularly the dynamical critical exponent $z$ and correlation length exponent $\nu$, characterize universal scaling behavior near quantum critical points.

\begin{figure}[htb]
  \centering
  \includegraphics[width=0.42\textwidth]{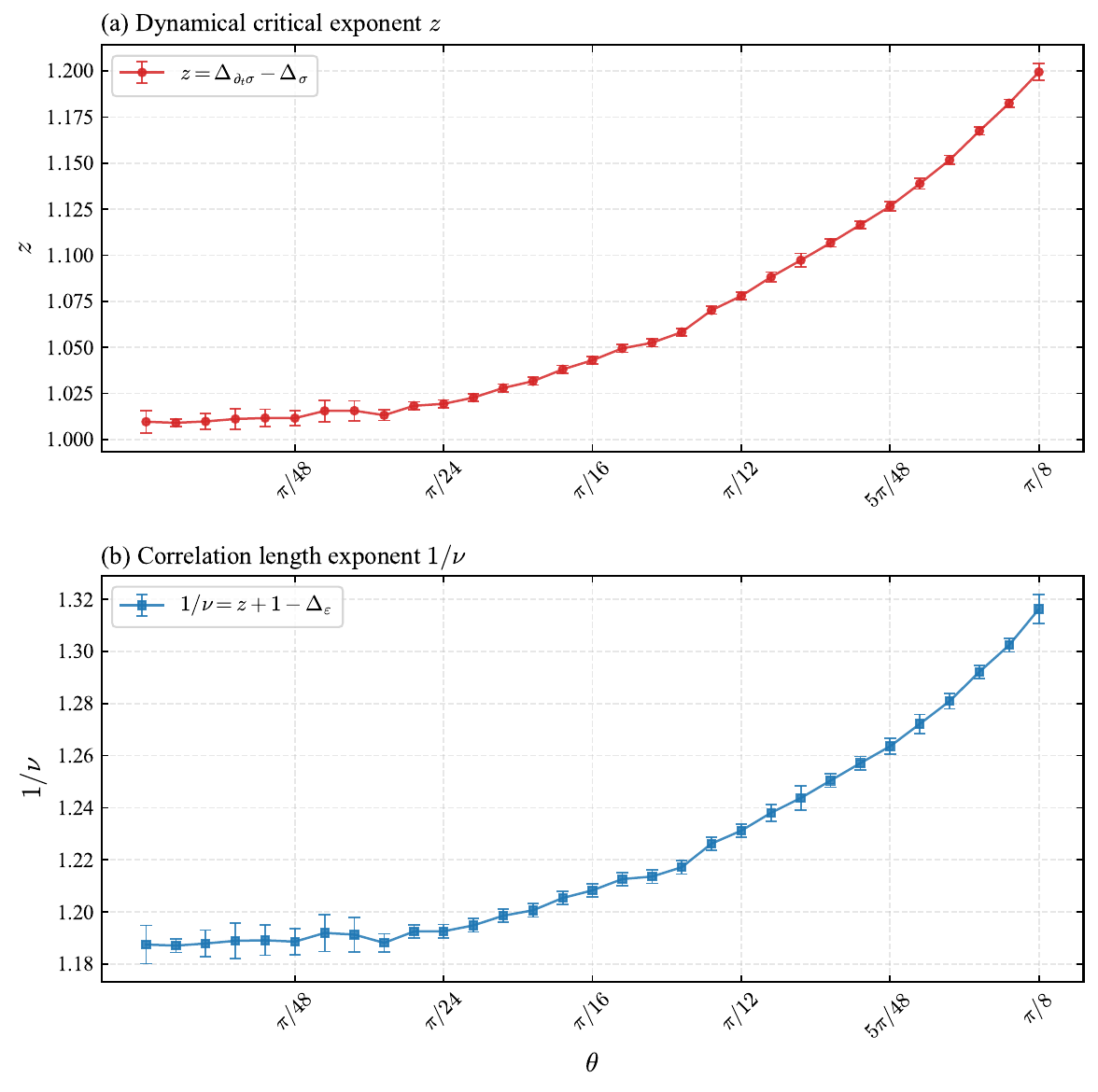}
  \caption{Critical exponents extracted from MERA analysis showing the emergence of anisotropic scaling. Both the dynamical exponent $z(\theta)$ and correlation length exponent $\nu(\theta)$ are displayed. Error bars represent statistical uncertainty only, estimated from internal-consistency residuals (see Appendix~\ref{subsec:error_analysis}). An additional $\theta$-independent systematic uncertainty of $\pm 0.010$ ($z$) and $\pm 0.024$ ($1/\nu$) from finite bond dimension is not shown.}
  \label{fig:critical-exponents-theta}
\end{figure}

The dynamical exponent $z$ exhibits distinct scaling regimes:
\begin{equation}
z(\theta) = \Delta_{\partial_t \sigma}(\theta) - \Delta_{\sigma}(\theta)
\end{equation}

For small chiral angles $\theta \lesssim \pi/16$, $z$ remains close to unity. For larger angles $\theta \gtrsim \pi/12$, $z$ increases more rapidly, reaching $z=1.199$ at $\theta=\pi/8$, signaling pronounced anisotropic scaling ($z\neq 1$).

The correlation length exponent follows the hyperscaling relation:
\begin{equation}
\frac{1}{\nu} = 1 + z - \Delta_\epsilon
\end{equation}

At $\theta = 0$, we recover $z = 1$ and $\nu = 5/6$, consistent with conformal invariance. For small $\theta$ values, $z$ shows modest deviations from unity (1.007 to 1.047), indicating weak anisotropic scaling. However, as $\theta$ increases beyond $\pi/12$, $z$ exhibits pronounced growth, demonstrating strong conformal symmetry breakdown. Correspondingly, the correlation length exponent $\nu$ shows systematic decrease with increasing $\theta$, with $1/\nu$ growing from 1.189 to 1.316, reflecting the enhanced role of fluctuations in the non-conformal regime.

\begin{table}[htb]
\centering
\footnotesize
\begin{tabular}{ccccc}
\hline
$\theta$ & $z$ (MERA) & $z$ (DMRG) & $1/\nu$ (MERA) & $1/\nu$ (DMRG) \\
\hline
$\pi/48$ & $1.012(4)$ & $1.00(7)$ & $1.189(5)$ & $1.20(9)$ \\
$\pi/24$ & $1.019(2)$ & $1.01(8)$ & $1.193(3)$ & $1.21(8)$ \\
$\pi/16$ & $1.043(2)$ & $1.02(1)$ & $1.208(3)$ & $1.22(3)$ \\
$\pi/12$ & $1.078(2)$ & $1.07(6)$ & $1.231(3)$ & $1.25(1)$ \\
$5\pi/48$ & $1.127(2)$ & $1.13(3)$ & $1.264(3)$ & $1.27(7)$ \\
$\pi/8$ & $1.199(5)$ & $1.22(7)$ & $1.316(6)$ & $1.32(4)$ \\
\hline
\end{tabular}
\caption{Comparison of critical exponents between MERA and DMRG methods. DMRG results from Ref.~\cite{Samajdar:2018xkm}. MERA uncertainties are statistical only (internal-consistency residuals); an additional $\theta$-independent systematic uncertainty of $\pm 0.010$ ($z$) and $\pm 0.024$ ($1/\nu$) from finite bond dimension applies to all rows.}

\label{tab:critical-exponents-comparison}
\end{table}

Our results are in good agreement with the DMRG calculations, validating the effectiveness of the MERA tensor network approach in non-conformal critical systems.

\subsubsection{Operator Product Expansion Coefficients}
\label{sec:ope-structure}

The operator product expansion coefficients encode fusion rules and correlation function structure of the critical theory.

\begin{figure}[htb]
    \centering
    \includegraphics[width=0.45\textwidth]{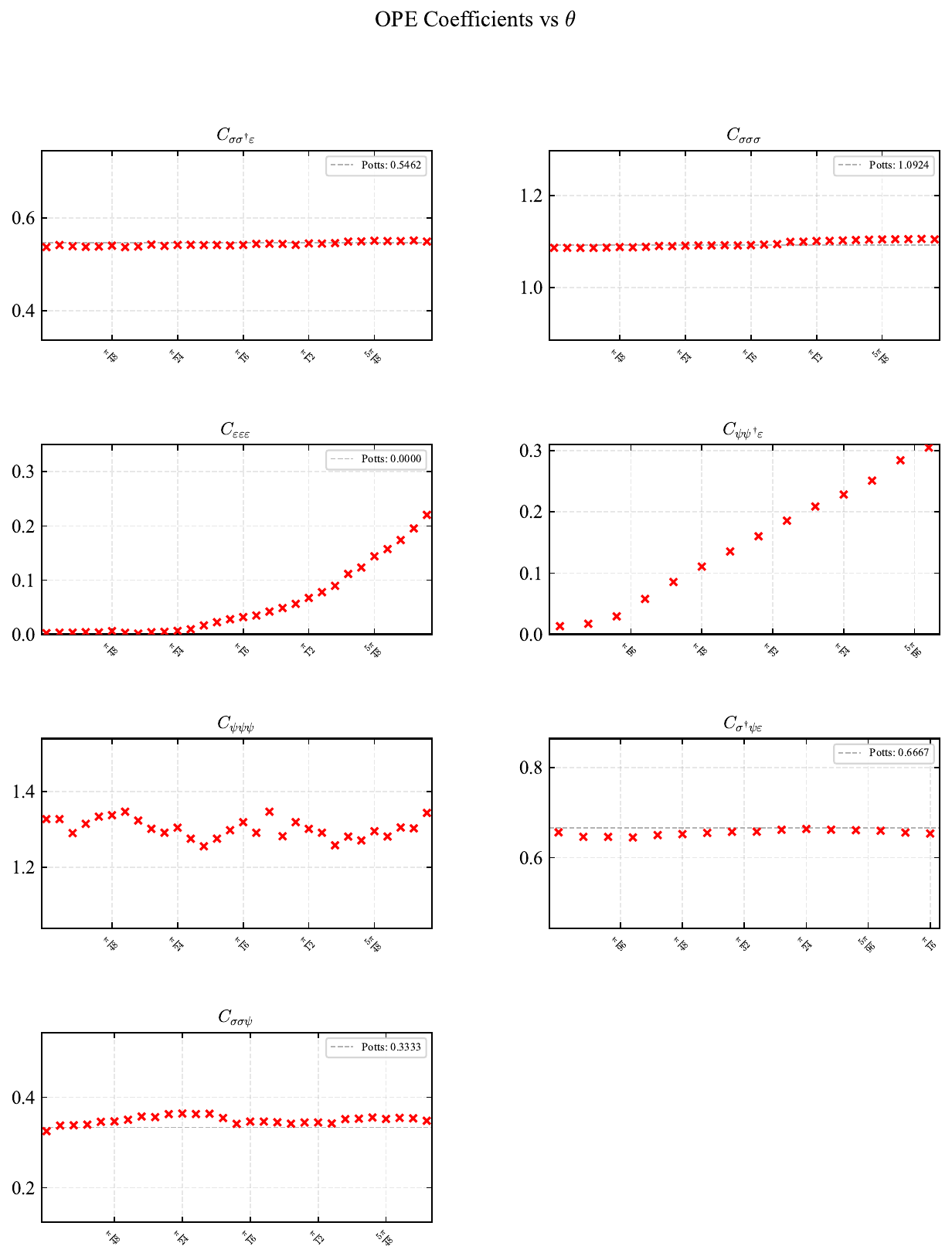}
    \caption{$\theta$-dependence of operator product expansion coefficients showing selective sensitivity to chiral deformation. Dashed horizontal lines indicate Potts theoretical values at $\theta=0$ where available.}
    \label{fig:ope-coefficients-theta}
\end{figure}

The OPE analysis reveals a rich structure of symmetry-protected and symmetry-broken coefficients. Normalization conditions fix the OPE coefficients $C_{\sigma\sigma^{\dagger}1}$, $C_{\psi\psi^{\dagger}1}$, and $C_{\epsilon\epsilon1}$ to unity by convention. The chiral $\mathbb{Z}_3$ symmetry imposes strict selection rules, causing most coefficients to vanish due to charge conservation and leaving only a subset of allowed three-point functions.

Among the non-trivial OPE coefficients, only $C_{\epsilon\epsilon\epsilon}$ and $C_{\psi\psi^{\dagger}\epsilon}$ exhibit pronounced dependence on $\theta$, reflecting their specific roles in chiral deformation. In contrast, the remaining coefficients $C_{\sigma\sigma^{\dagger}\epsilon}$, $C_{\sigma\sigma\sigma}$, $C_{\psi\psi\psi}$, $C_{\sigma^{\dagger}\psi\epsilon}$, and $C_{\sigma\sigma\psi}$ show remarkable stability, remaining essentially constant as $\theta$ varies. This stability suggests protection by the underlying symmetry structure, even as conformal invariance is broken.

The extraction of OPE coefficients represents a unique MERA capability, providing direct access to correlation function data challenging to obtain through other numerical methods. The selective $\theta$-dependence offers insights into symmetry breaking mechanisms and non-conformal scaling emergence.

It should be noted that for $z \neq 1$, the full OPE involves scaling functions $f_{ab}^{\;c}(u)$ of the invariant ratio $u = t/|x|^z$ (see \eqref{eq:anisotropic_OPE}). Our extracted coefficients $C_{ab}^{\;c} = f_{ab}^{\;c}(0)$ represent only the equal-time ($u=0$) slice of these functions. Consequently, the apparent stability of most OPE coefficients with $\theta$ may partly result from the equal-time projection, with additional $\theta$-dependence potentially residing in $f_{ab}^{\;c}(u \neq 0)$. Nevertheless, our data provide direct numerical constraints on $f_{ab}^{\;c}(0)$ even though the full functions remain undetermined. As discussed in \secref{sec:discussion}, the Heisenberg equation-of-motion method can systematically extract higher-order Taylor coefficients $f'^{\;c}_{ab}(0)$ from ground-state MERA, representing an important next step toward characterizing the complete anisotropic OPE structure.

\subsubsection{Effective Central Charge}
\label{sec:central-charge-results}

\label{sec:central-charge-method}
We define an effective central charge from entanglement entropy as
\begin{equation}
  c_{\mathrm{eff}} = 3\,\frac{S_2 - S_1}{\ln b},
  \label{eq:c-eff-mera}
\end{equation}
where $S_1$ and $S_2$ are the symmetrized one- and two-site entropies at the top layer and $b = 2$ is the scale ratio of the modified-binary MERA. For $z=1$ (CFT), $c_{\mathrm{eff}}$ reduces to the Virasoro central charge $c$, as defined by the Calabrese--Cardy formula $S(\ell) = \frac{c}{3}\ln\ell$~\cite{Calabrese:2004eu}. For $z \neq 1$, no Virasoro algebra exists, but the entanglement entropy still exhibits logarithmic scaling $S(\ell) \sim \kappa \ln\ell$, where $\kappa$ encodes the effective number of entangling degrees of freedom. Our $c_{\mathrm{eff}}$ thus measures the effective entanglement degrees of freedom at the MERA coarse-graining scale, rather than a true central charge.

\begin{figure}[t]
  \centering
  \includegraphics[width=0.38\textwidth]{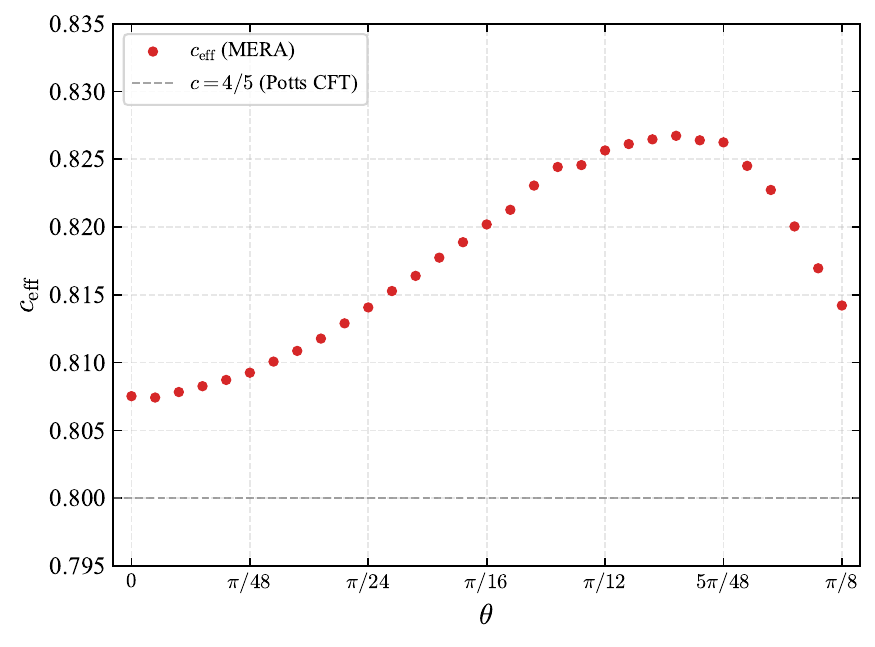} 
  \caption{Effective central charge $c_{\mathrm{eff}}$ versus the chiral angle $\theta$ along the critical line. The dashed line marks the Potts CFT value $c=4/5$.}
  \label{fig:central-charge-vs-theta}
\end{figure}

We evaluate $c_{\mathrm{eff}}$ along the critical line using the top-layer density matrix $\rho$ and the scale ratio $b = 2$. Figure~\ref{fig:central-charge-vs-theta} shows a smooth evolution with the chiral angle $\theta$.

At $\theta=0$ (Potts point), we obtain $c_{\mathrm{eff}}\approx 0.8075$, consistent with the expected CFT value $c=4/5$ within a small method-dependent bias. As $\theta$ increases, $c_{\mathrm{eff}}$ rises to a broad maximum around $\theta\approx \pi/10$ with $c_{\mathrm{eff}}\approx 0.8267$, and then gradually decreases to $\approx 0.814$ at $\theta=\pi/8$. This nonmonotonic, dome-like behavior can be understood as a competition between two effects: (i)~the chiral deformation enhances inter-operator entanglement, causing $c_{\mathrm{eff}}$ to rise; (ii)~at larger $\theta$, the growing dynamical exponent $z > 1$ compresses temporal correlations relative to spatial ones, reducing the spatial entanglement and causing $c_{\mathrm{eff}}$ to decrease. The peak at $\theta \approx \pi/10$ marks the approximate balance point between these two competing tendencies. The values remain close to the conformal benchmark across the line.

\subsection{Convergence Analysis and Numerical Precision}
\label{sec:convergence}

The reliability of our results depends on the convergence of the MERA optimization. We work directly in the thermodynamic limit using an infinite (scale-invariant) MERA, so there are no physical boundaries and hence no edge effects. We assess numerical precision by systematically varying the bond dimension $\chi$ and key optimization parameters. 

Bond-dimension convergence shows that $\chi=36$ is sufficient for all quantities reported here: scaling dimensions are stable within $\pm 10^{-3}$ and derived critical exponents within $\pm 10^{-2}$ upon increasing $\chi$ (e.g., from $\chi=32$ to $\chi=36$). Convergence of the variational energy density typically requires $10^3$--$10^4$ update steps (sweeps), after which the energy density and local reduced density matrices become stationary within a relative tolerance $\sim 10^{-8}$. 

We further check that the extracted scaling data are insensitive to the MERA architecture used to approach the fixed point, including the number of transitional layers and optimization schedules. In particular, varying the number of transitional layers (and other optimizer settings) changes scaling dimensions only within the quoted uncertainties, indicating that residual non-scale-invariant effects are below our error bars.

\subsection{Summary}
\label{sec:results-summary}

Our comprehensive MERA analysis establishes several key findings. From a methodological perspective, MERA provides precise access to universal scaling properties of non-conformal critical systems, with finite-$\chi$ systematic bias below $3\%$ and $\theta$-dependent statistical uncertainties below $0.5\%$, validating the tensor network approach for studying quantum criticality beyond the conformal paradigm. 

The chiral clock model exhibits smooth evolution of effective scaling exponents from conformal ($z = 1$) to anisotropic ($z \approx 1.2$) behavior. This smooth variation is consistent with both a continuous family of fixed points and a slow RG flow between the Potts and an anisotropic fixed point, as discussed in \secref{sec:discussion}. The complete scaling dimension spectrum reveals anisotropic tower structure emergence, with temporal and spatial descendants acquiring distinct scaling properties as the system deviates from conformal invariance. Furthermore, the fusion algebra exhibits selective sensitivity to chiral deformation, with some coefficients protected by discrete symmetry while others reflect conformal invariance breakdown, providing insights into the interplay between symmetry and criticality.

These results provide comprehensive characterization of non-conformal scaling in the chiral clock model, establishing MERA as a powerful tool for investigating quantum criticality beyond the conformal paradigm.

\section{Discussion}
\label{sec:discussion}

In this work, we have demonstrated the application of MERA tensor networks to study the critical behavior of the $\mathbb{Z}_3$ chiral clock model, particularly focusing on a line of quantum phase transitions with nontrivial dynamical critical exponent. Our results provide compelling evidence that MERA can serve as a powerful tool for extracting universal scaling information from critical systems that lie beyond the conformal field theory paradigm.

On the physics side, our results were consistent with a set of critical exponents that vary continuously along the critical line. Previous DMRG studies~\cite{Samajdar:2018xkm,Zhuang:2015fpa} have obtained similar results, although this represents an unusual scenario from a field theory perspective. A more standard situation would have just two important fixed points, the 3-state Potts point and an anisotropic fixed point at larger $\theta$. We could not explicitly verify this scenario since the scaling dimension data around dimension $1+z$ became too noisy to reliably detect the existence of a relevant operator causing a flow along $\theta$. It would be very interesting to return to this issue in the future, perhaps incorporating more of the duality structure of the model into the study \cite{Cardy_1993,Mong_2014} or working perturbatively around the Potts point to study the RG flows.

How might this addressed within MERA? Our approach is in essence treating an infinite system with a four layer MERA network. We find empirically that while the first and second layers are distinct, the third and fourth layers are much closer to each other. One interpretation is that the first few layers form what are sometimes called transitional layers, which then give way to the emergence of scale invariance in which all subsequent layers are approximately equal. In our case, we extract the scaling data from the approximately scale invariant layers, leading to the results presented in this paper. How does this relate to the slow flow hypothesis? Using the perturbative results of Cardy \cite{Cardy_1993}, the perturbing operator dimension $\Delta = 9/5$ is close to the marginal value $1+z=2$ (at the Potts point), with a $\beta$-function coefficient of only $1/5$. The crossover length obeys a power law, $\ell^{*}\;\sim\;\theta^{-5}$. For $\theta=\pi/8$ this gives $\ell^{*}\sim 10^{2}$ lattice sites, while for $\theta=\pi/48$ it yields $\ell^{*}\sim 10^{6}$. Across our entire parameter range, $\ell^{*}$ exceeds the length scales covered by the transitional layers, so the scale-invariant MERA tensor represents an intermediate point along the RG trajectory rather than the true infrared fixed point. MERA yielding smoothly varying effective exponents is therefore self-consistent with the slow-flow scenario. These effective exponents carry physical meaning: they describe the system's actual scaling behavior at the scales accessible to MERA and, notably, at the system sizes relevant to Rydberg atom experiments ($L\sim 50$--$250$). In principle, comparing MERA results obtained with different numbers of transitional layers could probe the RG flow rate directly: if the effective exponents shift with the number of transitional layers, this would provide further support for the slow-flow hypothesis.

On the methodological side, the reliability of our MERA method is underpinned by three elements: rigorous convergence diagnostics; stability of scaling data across tested bond dimensions and network depth; and cross-validation against CFT benchmarks (Potts point) and with DMRG critical exponents from the literature. These controls produce consistent operator spectra and critical exponents across $\theta$. Full procedures and uncertainty handling are provided in the Methods and Appendix. Continuing to improve the methods is another target for future work.

In particular, the computational cost of MERA optimizations limits the bond dimension, potentially restricting the range of the computable scaling spectrum and affecting the accuracy of extracted scaling data. Furthermore, the connection between the numerically obtained operator spectrum and the underlying field theory structure requires further theoretical development for these non-relativistic field theories.

We also see several promising avenues for extending this work. First, the natural generalization to other $\mathbb{Z}_N$ chiral clock models with $N>3$ offers opportunities to explore how the richness of non-conformal behavior scales with the local Hilbert-space dimension and to investigate potential universality classes beyond the present study. Second, the extension to higher spatial dimensions represents a significant challenge, as two-dimensional chiral systems may exhibit fundamentally different critical behavior, including potential topological phase transitions and anyonic excitations that are absent in one dimension. Third, an important direction is to systematically characterize the non-conformal scaling functions $\mathcal{F}_{ij}\!\left(\frac{t_{12}}{|x_{12}|^{z}}\right)$ appearing in dynamical correlation functions, which capture anisotropic space--time dependence beyond conformality and encode operator mixing and selection rules. Concretely, one can access the short-time expansion of $\mathcal{F}_{ij}$ directly from ground-state MERA computations, without performing explicit real-time evolution: in the Heisenberg picture, $\partial_t O_i(t)=i[H,O_i(t)]$, implying the equation-of-motion identity
\begin{equation}
\left.\partial_t \langle O_i(x,t)\,O_j(0,0)\rangle\right|_{t=0}
= i\,\langle [H,O_i(x)]\,O_j(0)\rangle ,
\end{equation}
and higher nested commutators $\langle \mathrm{ad}_H^n(O_i(x))\,O_j(0)\rangle$ determine higher derivatives $\mathcal{F}^{(n)}_{ij}(0)$. This provides a systematic and purely lattice-based route to dynamical information at non-conformal fixed points.

From a methodological perspective, several technical advances would significantly enhance the scope of MERA-based studies of non-conformal criticality. The development of more efficient tensor-optimization algorithms, potentially leveraging machine learning techniques or automatic differentiation, could enable access to larger bond dimensions and improved numerical accuracy. Additionally, the systematic incorporation of global on-site and lattice symmetries into the MERA structure, along the lines of symmetry-adapted tensor-network and MERA constructions~\cite{PhysRevB.88.121108,WEICHSELBAUM20122972,PhysRevB.99.195139}, could reduce computational costs by promoting block-sparse tensors while providing direct access to quantum numbers and selection rules. The extension of MERA to finite-temperature studies would open connections to experimental realizations in quantum simulators, where thermal effects are often unavoidable~\cite{PhysRevB.94.085101,PhysRevLett.115.200401}.

Finally, the methodology established here provides a pathway for exploring other classes of non-conformal critical phenomena. One example is transitions in driven-dissipative systems where the non-equilibrium nature fundamentally alters the critical properties. The systematic extraction of operator spectra and OPE coefficients demonstrated in this work could prove essential for characterizing such exotic critical points, where traditional field-theoretical approaches are often inadequate. The foundation laid here thus opens new frontiers for understanding quantum criticality beyond the conformal paradigm.

\textit{Acknowledgements:} SG and BS acknowledge support from the US Department of Energy QuantISED 2.0 program as part of the GeoFlow consortium under grant number DE-SC0019380. We thank Glen Evenbly, Yifan Wang, and Soonwon Choi for discussions.

\bibliographystyle{apsrev4-2}
\bibliography{bibliography}

\appendix
\section{Appendix}
\label{sec:appendix}

\subsection{Details of MERA implementation}
\label{subsec:mera_details}

This subsection summarizes the concrete numerical protocols used in our calculations. Notation: $u$ denotes the disentangler, $\omega$ and $v$ the isometries, $\chi$ the bond dimension, and $(\rho_{AB},\rho_{BA})$ the top-layer density matrices used for warm starts.

\paragraph*{Convergence criterion at fixed $(\theta, f_\ast)$.}
For every critical point $(\theta, f_\ast)$ the MERA optimization is declared converged when the following two conditions are simultaneously satisfied between successive checkpoints (checkpoints are recorded every 200 optimizer iterations):
\begin{enumerate}
  \item The relative change of the lowest ten scaling dimensions (excluding the identity) is $\le 0.5\%$;
  \item The relative change of a predefined set of primary OPE coefficients is $< 1\%$.
\end{enumerate}
Here “relative change” refers to $|x^{(t)}-x^{(t-1)}|/|x^{(t)}|$ for the tracked quantity $x$ at two consecutive checkpoints (indexed by $t$). The monitored OPE set consists of the lowest-lying channels used in the main text.

\paragraph*{Progressive bond-dimension ramp at the first $\boldsymbol{\theta}$.}
For the first $\theta$ value we adopt a progressive increase of the bond dimension $\chi$ to reach high accuracy in a stable manner. We first optimize at a smaller $\chi$ until the above convergence test is met, obtaining tensors $(u,\omega,v)$ and the top-layer density matrices. These objects are then \emph{embedded} into a larger bond dimension $\chi'\!>\!\chi$ as follows: (i) copy the converged subblocks to the upper-left corner of the new tensors, (ii) initialize the added degrees of freedom to (near-)zero and re-orthonormalize to satisfy the unitary/isometry constraints (polar/QR projection), and (iii) expand optimizer state (e.g., momenta) by padding zeros in the new indices. The top density matrices are lifted by block-diagonal padding and trace renormalization. We then continue optimization at $\chi'$ from this warm start. In this work we use $\chi\in\{12,16,24,36\}$.

\paragraph*{Adiabatic crawling along the critical line.}
Once a critical point is converged, subsequent points are obtained by adiabatic crawling: for the next $(\theta, f_\ast(\theta))$, we load the converged $(u,\omega,v)$ and top density matrices $(\rho_{AB},\rho_{BA})$ from the nearest previously-solved point (typically the one with the closest larger $\theta$) and resume optimization under the current Hamiltonian. To suppress long-time drift we perform periodic cold restarts at fixed anchor spacing in $\theta$ (we use $\Delta\theta=\pi/48$) and, by default, carry out crawling at the largest $\chi$.

\paragraph*{Avoiding poor local minima.}
We combine three complementary strategies to mitigate trapping:
\begin{itemize}
  \item \emph{Isometric parametrization with manifold retraction.} All updates respect the unitary/isometry constraints via SVD/polar (or Cayley/EV) retractions, keeping tensors close to a Stiefel manifold and reducing gauge redundancy/nonconvex pathologies (cf.~Ref.~\cite{Geng:2021msw}).
  \item \emph{Randomly mixed optimizers.} We alternate an EV-style update with geometry-aware SGD steps (SVD/Cayley variants), which provide diverse curvature response and momentum, improving escape from shallow basins.
  \item \emph{Restarts and multi-start.} When energy or tracked observables stagnate, or convergence diagnostics (spectrum/OPE consistency) fail, we trigger restarts and run multiple random seeds in parallel, retaining the best solution.
\end{itemize}

The above procedures were used throughout all reported parameter points.

\subsection{Optimization Algorithm}
\label{subsec:optimization}

The optimization of MERA tensors is realized as a fully differentiable program implemented in \textsc{PyTorch}. Each training epoch begins with a short Evenbly–Vidal sweep that supplies a low-energy seed, after which all tensors are refined simultaneously by Riemannian gradient descent on the Stiefel manifold following the framework of~\cite{Geng:2021msw}.  Denoting by $X$ a flattened tensor and by $dX=\partial_{X}E$ its Euclidean energy gradient, the projected tangent-space direction is
\[
  G = dX - \tfrac12\bigl(X X^{\dagger} dX + X dX^{\dagger} X\bigr),
\]
and the update $X\rightarrow X_{\text{new}}$ is obtained through either an SVD retraction $X_{\text{new}}=UV^{\dagger}$ with $(X-\alpha G)=U\Sigma V^{\dagger}$ or, for reduced cost, a Cayley transform.  Reverse-mode automatic differentiation constructs the adjoint computation graph on the fly, providing exact Riemannian gradients at less than twice the cost of the forward pass and obviating the need for hand-derived Jacobians.  In practice we alternate between stochastic gradient descent and Adam steps on the manifold, interleaved with occasional Evenbly–Vidal updates, a strategy that mitigates shallow local minima and accelerates convergence to machine precision ($\sim10^{-10}$) within a few hundred iterations.

All contractions are executed on GPUs with mixed-precision arithmetic; batched kernels and gradient checkpointing confine the memory footprint to $\mathcal{O}(\chi^{6})$, enabling bond dimensions up to $\chi\approx50$ on a single high-end device.  The resulting hybrid differentiable-programming workflow unifies several optimizers in a common infrastructure, facilitates the inclusion of auxiliary loss terms such as spectral regularisers, and permits back-propagation through hyper-parameters like the coarse-graining depth.

\subsection{Error Analysis and Statistical Reliability}
\label{subsec:error_analysis}

We quantify uncertainties via a pragmatic, three-pronged robustness analysis tailored to MERA: (i) a $\theta{=}0$ conformal benchmark that sets a baseline accuracy; (ii) internal consistency diagnostics that check exact relations and symmetry-implied degeneracies; and (iii) statistical variability estimated from repeated optimizations at representative points. Together these provide error bars for scaling dimensions, which we then propagate in the usual way to derived exponents (e.g., $z$ and $1/\nu$). We report conservative uncertainties by taking the maximum across the three methods when combining them. For derived quantities such as $z$ and $1/\nu$, where finite-$\chi$ biases partially cancel in differences of scaling dimensions, we adopt a separate treatment that distinguishes statistical from systematic uncertainties, as detailed in \S\ref{subsubsec:combined_errors}.

\subsubsection{Method 1: Theoretical Benchmark Analysis ($\theta = 0$)}
\label{subsubsec:theoretical_benchmark}

For the conformal limit $\theta = 0$, we validate our MERA implementation against the exactly known results of the critical 3-state Potts model.

According to Fig.~\ref{fig:scaling-dimensions-evolution}, our MERA calculations at $\theta = 0$ show excellent agreement with theoretical predictions, with relative errors less than 3\% for all quantities.

While the excellent agreement (relative errors $< 3\%$) at $\theta = 0$ establishes confidence in our MERA implementation for the integrable, conformal case, this does not imply that similar accuracy is maintained for $\theta \neq 0$. In the non-integrable, non-conformal regime, systematic errors may be significantly larger due to the breakdown of conformal symmetry and the increased complexity of the critical theory. The $\theta = 0$ benchmark primarily serves to validate our numerical implementation rather than to extrapolate error estimates across the entire parameter space.

\subsubsection{Method 2: Internal Consistency Diagnostics}
\label{subsubsec:hyperscaling}

We use two exact internal checks to diagnose systematics: (i) the spatial-derivative identity $\Delta_{\partial_x \Phi}-\Delta_{\Phi}=1$ for primary fields, and (ii) the degeneracy of symmetry-related conjugate pairs. Deviations (“residuals”) from these relations provide clean, model-independent diagnostics of systematic uncertainty.

Beyond the derivative identity, we also require degeneracy of conjugate pairs (e.g., $\sigma$ vs $\sigma^{\dagger}$, $\partial_x\sigma$ vs $\partial_x\sigma^{\dagger}$, $\partial_t\sigma$ vs $\partial_t\sigma^{\dagger}$, and $\psi$ vs $\psi^{\dagger}$). The corresponding residuals are combined into a single conservative systematic-uncertainty estimate per scaling dimension, which we then propagate to exponents.

The hyperscaling residual analysis provides a comprehensive diagnostic tool that captures systematic uncertainties across the parameter space, regardless of whether they originate from finite bond dimension truncation effects, optimization convergence limitations, or the intrinsic complexity of the non-conformal critical theory itself. The systematic analysis demonstrates that our MERA calculations maintain reasonable accuracy ($\sigma_{\mathrm{stat},z} \lesssim 0.005$) across most of the parameter range, with systematic error estimates that appropriately reflect the increasing theoretical complexity as the system deviates further from conformal invariance.

\subsubsection{Method 3: Statistical Sampling Analysis}
\label{subsubsec:statistical_sampling}

To assess reproducibility and optimiser-induced variability, we perform multiple independent MERA optimizations at a representative non-conformal point (e.g., $\theta=\pi/8$) under identical hyperparameters. Primary fields show the smallest spread, derivatives are moderately larger, and parafermions are the most sensitive. These spreads define a “statistical” uncertainty that we combine with the other methods. Overall, this triad of checks separates random optimization noise from systematic effects and yields robust, conservative error bars reported in Table~\ref{tab:final_errors}.

\subsubsection{Combined Error Analysis}
\label{subsubsec:combined_errors}

\begin{table}[t]
  \centering
  \begin{tabular}{cccccc}
  \hline
  $\theta$ & $\Delta_{\sigma}$ & $\Delta_{\epsilon}$ & $\Delta_{\psi}$ & $z$ & $1/\nu$ \\
  \hline
    $\pi/48$ & $0.1346(29)$ & $0.823(22)$ & $1.332(14)$ & $1.012(4)$ & $1.189(5)$ \\
    $\pi/24$ & $0.1316(16)$ & $0.827(23)$ & $1.346(14)$ & $1.019(2)$ & $1.193(3)$ \\
    $\pi/16$ & $0.1256(17)$ & $0.835(25)$ & $1.371(14)$ & $1.043(2)$ & $1.208(3)$ \\
    $\pi/12$ & $0.1180(18)$ & $0.847(27)$ & $1.403(14)$ & $1.078(2)$ & $1.231(3)$ \\
    $5\pi/48$ & $0.1078(20)$ & $0.863(30)$ & $1.452(14)$ & $1.127(2)$ & $1.264(3)$ \\
    $\pi/8$ & $0.0944(32)$ & $0.883(33)$ & $1.531(14)$ & $1.199(5)$ & $1.316(6)$ \\
  \hline
  \end{tabular}
  \caption{Representative scaling dimensions and critical exponents. Uncertainties on $\Delta_\sigma$, $\Delta_\epsilon$, and $\Delta_\psi$ include the finite-$\chi$ Potts-benchmark bias (combined via the maximum of all three methods). Uncertainties on $z$ and $1/\nu$ are statistical only (internal-consistency residuals); an additional $\theta$-independent systematic uncertainty of $\pm 0.010$ ($z$) and $\pm 0.024$ ($1/\nu$) applies to all rows.}
  \label{tab:final_errors}
  \end{table}

For scaling dimensions ($\Delta_\sigma$, $\Delta_\epsilon$, $\Delta_\psi$), we combine uncertainties from all three methods using a conservative maximum rule, as these absolute values carry the full finite-$\chi$ bias. For derived quantities ($z$ and $1/\nu$), which involve differences of scaling dimensions where the finite-$\chi$ bias partially cancels, we separate the uncertainty into a $\theta$-dependent statistical component (from internal-consistency residuals) and a $\theta$-independent systematic component (propagated from Potts-benchmark biases: $\sigma_{\mathrm{sys},z}=|b_{\Delta_{\partial_t\sigma}}-b_{\Delta_\sigma}|=0.010$, $\sigma_{\mathrm{sys},1/\nu}=\sqrt{\sigma_{\mathrm{sys},z}^2+b_{\Delta_\epsilon}^2}=0.024$). This separation follows standard practice in precision measurements, where systematic offsets common to all data points are reported separately from point-to-point statistical fluctuations.

Table~\ref{tab:final_errors} presents representative uncertainties in our data, including all the primaries and the critical exponents.

\end{document}